\documentclass[twocolumn]{aastex63}

\newcommand{\degree}{\mbox{$^{\circ}$}}
\usepackage{totcount}
\newtotcounter{citnum} 
\def\oldbibitem{} \let\oldbibitem=\bibitem
\def\bibitem{\stepcounter{citnum}\oldbibitem}
\usepackage{floatrow}
\usepackage[caption=false]{subfig}
\accepted{September, 19, 2020}
\submitjournal{AJ}

\shorttitle{TESS Observations of the Hot Jupiter Exoplanet XO-6b}
\shortauthors{Ridden-Harper, Turner \& Jayawardhana}
\graphicspath{{./}}

\begin{document}

\title{TESS Observations of the Hot Jupiter Exoplanet XO-6b: No Evidence of  Transit Timing Variations}

\correspondingauthor{Andrew Ridden-Harper}
\email{arr224@cornell.edu}

\correspondingauthor{Jake D. Turner}
\email{astrojaketurner@gmail.com, jaketurner@cornell.edu}

\author[0000-0002-5425-2655]{Andrew Ridden-Harper}
\affil{Department of Astronomy and Carl Sagan Institute, Cornell University, Ithaca, New York 14853, USA}

\author[0000-0001-7836-1787]{Jake D. Turner}
\affil{Department of Astronomy and Carl Sagan Institute, Cornell University, Ithaca, New York 14853, USA}

\author{Ray Jayawardhana}
\affil{Department of Astronomy, Cornell University, Ithaca, New York 14853, USA}

\newcommand\SigmaRuleOut{10}
\newcommand\NewPeriod{3.7649893$\pm$0.0000037 days}
\newcommand\NewTransitEpoch{2456652.7157$\pm$0.0022 BJD$_{TDB}$}

\newcommand\TessTTVlim{2.5 minutes}



\begin{abstract}
From previous ground-based observations, the hot Jupiter exoplanet XO-6b was reported to exhibit apparently periodic transit timing variations (TTVs), with a semi-amplitude of 14 minutes and a period of about 450 days. These variations were interpreted as being due to a resonant perturbation between XO-6b and a hitherto unknown low-mass planet orbiting the same star. To understand this enigmatic planetary system better, we analysed three sectors of data, spanning over seven months, from the Transiting Exoplanet Survey Satellite (TESS), which produces high-quality light curves that are well suited to characterizing exoplanets and searching for TTVs. Here we present an updated orbital period of \NewPeriod{} and a transit epoch of \NewTransitEpoch{}. The planetary parameters we report, while consistent with their discovery values, have greatly improved precision. Notably, we find no evidence for TTVs:  we can rule out TTVs $\gtrsim$ \TessTTVlim{} at the 3$\sigma$ level.  Therefore, the TESS data have sufficient precision and time baseline to reveal readily the previously reported TTVs of approximately 10 minutes. Our findings highlight TESS's capabilities for robust follow-up, and confirm that TTVs are rarely seen in hot Jupiters, unlike is the case with small planets.   

\end{abstract}

\keywords{planets and satellites: dynamical evolution and stability --- planets and satellites: gaseous planets ---  planet–star interactions ---  planets and satellites: individual (XO-6b) --- methods: observational --- techniques: photometric}


\section{Introduction} \label{sec:intro}

Gravitational interactions between bodies in a planetary system can cause a transiting exoplanet's time of transit to vary. Analysis of such transit timing variations (TTVs) can reveal important dynamical insights into a planetary system.  About 130 small planets were found to exhibit TTVs from \textit{Kepler} data \citep{Mazeh2013}. However, TTVs have only been observed in a handful of hot Jupiters. For example, a variation in the transit time and impact parameter of Kepler-13Ab has revealed evidence for spin-orbit precession caused by its host star's rapidly rotating quadrupole moment \citep[e.g.,][]{Masuda2015, Herman2018, Szabo2020}, allowing the stellar surface to be mapped \citep{Szabo2014}. \citet{Becker2015} found that WASP-47b exhibits sinusoidal TTVs caused by two smaller short-period planets in the same system. WASP-12b also exhibits TTVs, presumably resulting from orbital decay or apsidal precession \citep{Maciejewski2016, Patra2017, Maciejewski2018}; recently \citet{Yee2020} offered new evidence that favors tidally-induced orbital decay as the explanation. 

The Transiting Exoplanet Survey Satellite (TESS) \citep{Ricker2014} produces high-quality data that are well suited to searching for TTVs (e.g. \citealt{Hadden2019,Pearson2019}). Using TESS data, \citet{Bouma2019} found that the transits of WASP-4b occurred about 82 seconds earlier than expected, and determined that its 1.3-day orbital period is decreasing at a rate of about $-12.6$ $\pm$ 1.2 ms per year. \citet{Southworth2019} confirmed the presence of TTVs, but revised the orbital period decay rate to $-9.2$ $\pm$ 1.1 ms per year. More recently, a homogeneous analysis of 124 transits of WASP-4b, observed with several different telescopes,  found that its rate of change in orbital period is about half that found in previous studies \citep{Baluev2019,Baluev2020}.  Finally, \citet{Bouma2020} found that its orbital period changes by -8.64 ms per year. While it has been suggested that these TTVs may arise from orbital decay or apsidal precession \citep{Southworth2019}, recent findings indicate that they are due to the system accelerating towards the Sun at a rate of -0.0422  ms$^{-1}$ day$^{-1}$ \citep{Bouma2020}.

Here we focus on the hot Jupiter XO-6b, with a mass and radius of 1.9 R$_{Jup}$ and 2.07 M$_{Jup}$, respectively. It orbits a fast-rotating ($v sin i$ = 48 kms$^{-1}$), bright (V = 10.25 mag), hot ($T_{eff}$ = 6720 K) star and has an orbital period of 3.8 days \citep{Crouzet2017}. To characterize XO-6b better, \cite{Garai2020} observed its transits with telescopes at the Astronomical Institute of the Slovak Academy of Sciences and downloaded transit light curves from the Exoplanet Transit Database (ETD)\footnote{\url{http://var2.astro.cz/ETD/}} \citep{Poddany2010}. From these data, they reported that it exhibited intriguing periodic TTVs with a semi-amplitude of 14 minutes and period of about 450 days.  By fitting these transit timing variations with the publicly available TTV analysis package, \texttt{OCFit}\footnote{\url{https://github.com/pavolgaj/OCFit}} \citep{Gajdos2019OCFitRef}, they determined the two most plausible explanations to be 1) light-time effects (LiTE) due to a third unknown stellar-mass object in the XO-6 system, or 2) resonant perturbations between XO-6b and an unknown low-mass planet in the system. However, they found no evidence for a stellar mass object in radial velocity (RV) follow-up, and simultaneous fits to their transit timing and RV measurements did not yield a consistent solution; so they favor the second interpretation. If the latter interpretation were correct, the XO-6b system would resemble the recently discovered TOI-216 system which contains a pair of warm, large exoplanets that exhibit planet-planet interactions \citep{Dawson2019,Dawson2020}.

Motivated by the intriguing TTVs of XO-6b reported by \cite{Garai2020}, we investigated this system further by analysing its TESS light curves. Our paper is structured as follows.  Section \ref{sec:data} describes our data reduction method. Section \ref{sec:analysis} shows our analysis and Section \ref{sec:results} presents and discusses our results. Finally, Section \ref{sec:conclusion} offers our conclusions. 

\section{Observations} \label{sec:data}

XO-6b was observed by TESS in Sector 19 (November 27, 2019 to December 24, 2019), Sector 20 (December 24, 2019 to January 21, 2020) and Sector 26 (June 8, 2020 to July 4, 2020). These observations were processed by the Science Processing Operations Center (SPOC) pipeline, which produces light curves corrected for systematics and searches for transiting planets \citep{Jenkins2016}.  All of the data products produced by SPOC are publicly available from the Mikulski Archive for Space Telescopes (MAST)\footnote{\url{https://archive.stsci.edu/}}. We downloaded all of the data products for XO-6b, including the light curve (LC) files, data validation timeseries (DVT) files, and target pixel files (TPFs). The Presearch Data Conditioning (PDC) component of the SPOC pipeline corrects the light curves for pointing or focus related instrumental signatures, discontinuities resulting from radiation events in the CCD detectors, outliers, and flux contamination \citep{Jenkins2016}.  The light curve resulting from the PDC corrections is recorded as the PDCSAP\_FLUX, and was one of the data products considered in our analysis.  The PDCSAP\_FLUX is further processed by using a running median filter to remove any long-period systematics before the SPOC pipeline searches for transits\footnote{\url{https://exoplanetarchive.ipac.caltech.edu/docs/DVSummaryPageCompanion.html}}.  The length of the running median filter that was used is recorded in the file headers and in the case of the XO-6b data was 14.8, 15.2, and 15.0  hours in Sectors 19, 20 and 26, respectively. These light curves are recorded in the DVT file as LC\_INIT and were also considered in our analysis. 

For comparison, we also considered light curves that we produced from the TPFs using aperture photometry.  To do this, we followed the documentation\footnote{\url{https://exoplanet.readthedocs.io/en/v0.1.6/tutorials/tess/}} accompanying the \texttt{exoplanet} package \citep{exoplanet}.  We selected the optimal aperture that minimized the windowed scatter and detrended the resulting light curve with the pixel-level deconvolution (PLD) method used by the \texttt{Everest} package \citep{Luger2016}.  

\begin{figure*}
\plotone{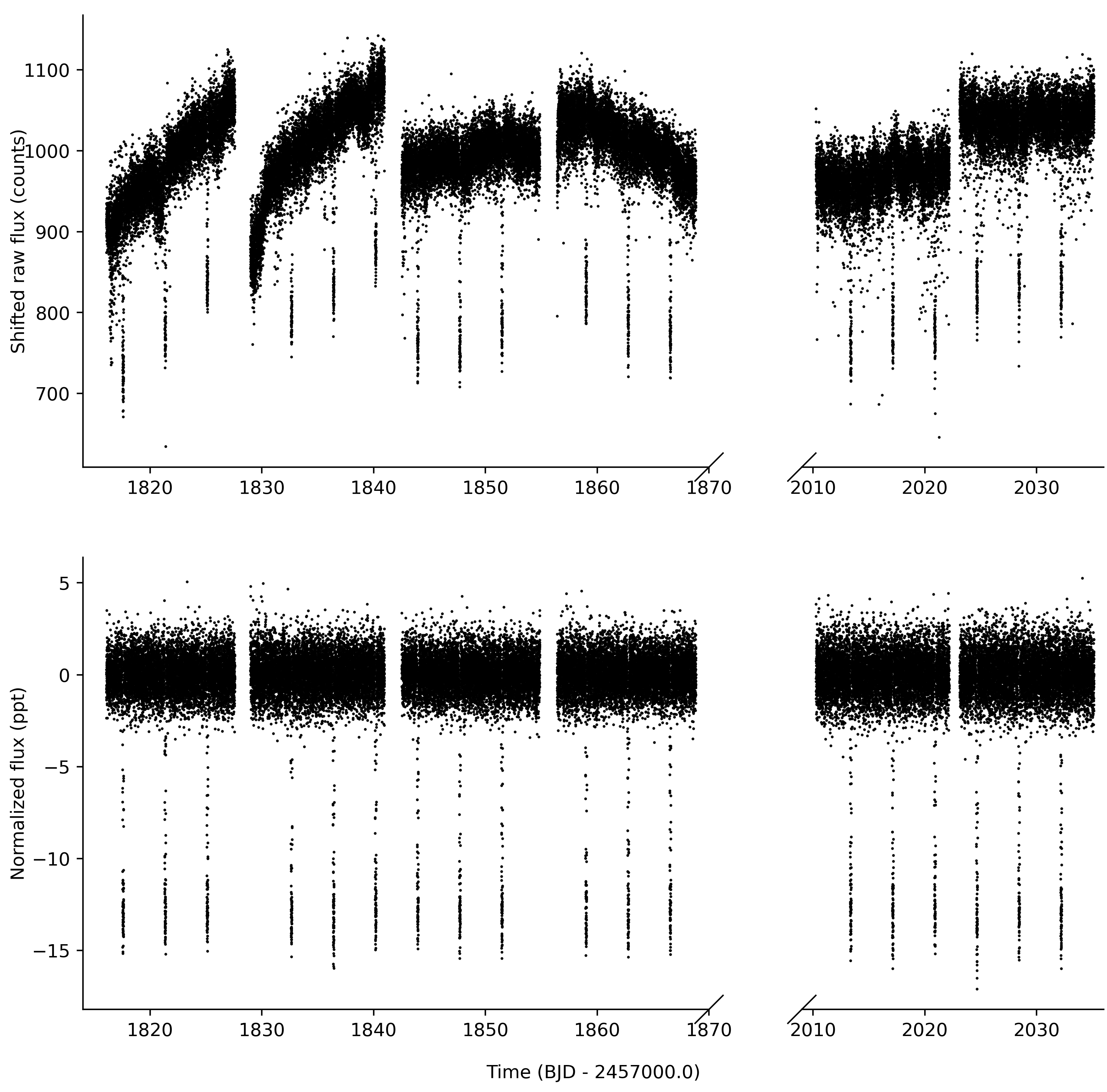}
\caption{TESS light curve of XO-6b in Sectors 19, 20 and 26. Top: Raw simple aperture photometry light curves, shifted vertically for clarity. Bottom: Detrended Data Validation Timeseries (DVT). }
\label{fig:lightcurves}
\end{figure*}

Of the three light curve detrendings that we considered, the DVT light curve had the least scatter, with a standard deviation on the out-of-transit baseline of 1.08 ppt, which was 3\% and 13\% lower than that of our light curves from the TPFs and the PDCSAP\_FLUX, respectively.  Therefore, we focused on the DVT light curves in our analysis. Nonetheless, we also analysed the transit timings of the other light curves and found them to be practically identical to what was derived from the DVT light curves (see Appendix \ref{app:diff_detrend}; Fig. \ref{fig:LC_comp}).  The raw light curve produced by the SPOC and detrended DVT light curve used in this analysis are shown in Figure \ref{fig:lightcurves}.

\section{Data analysis} \label{sec:analysis}

To find the best-fit to the TESS transits we use the EXOplanet MOdeling Package (\texttt{EXOMOP}; \citealt{Pearson2014,Turner2016,Turner2017})\footnote{\url{EXOMOPv7.0}; \href{https://github.com/astrojake/EXOMOP}{https://github.com/astrojake/EXOMOP} }, which generates a model transit using the analytic equations of \cite{Mandel2002}. \texttt{EXOMOP} uses a Differential Evolution Markov Chain Monte Carlo (DE-MCMC; \citealt{terBrrak2002}; \citealt{Eastman2013}) analysis to model the data and incorporates three different red noise methods (time-averaging, \citealt{Pont2006}; residual permutation, \citealt{Southworth2008}; wavelet, \citealt{Carter2009}) to assess and account for red noise in the light curve. A thorough description of \texttt{EXOMOP} can be found in \citet{Pearson2014} and \citet{Turner2016}.

We model each transit event in the TESS data (lower panel of Figure \ref{fig:lightcurves}) independently. Each transit is modeled with 20 chains and 20$^{6}$ links for the DE-MCMC model and we use the Gelman-Rubin statistic (\citealt{Gelman1992}) to ensure chain convergence (\citealt{Ford2006}). For each transit, the mid-transit time ($T_{c}$), planet-to-star radius ($R_{p }/R_{*}$), scaled semi-major axis ($a/R_{*}$), and inclination ($i$) are set as free parameters. The eccentricity, argument of periastron, period ($P_{orb}$), and linear and quadratic limb darkening coefficients are fixed during the analysis. The 
linear and quadratic limb darkening coefficients are taken from \citet{Claret2017} and are equal to 0.3158 and 0.2206, respectively. The light curve parameters derived for each individual transit can be found in Table \ref{tb:lighcurve_model} and the individual modeled light curves can be found in Figures \ref{fig:ind_transits_Sec19} --\ref{fig:ind_transits_sec26} in Appendix \ref{app:individual_transits}. All parameters for each transit event are consistent within 1$\sigma$ of every other transit. 

\begin{figure*}
\plotone{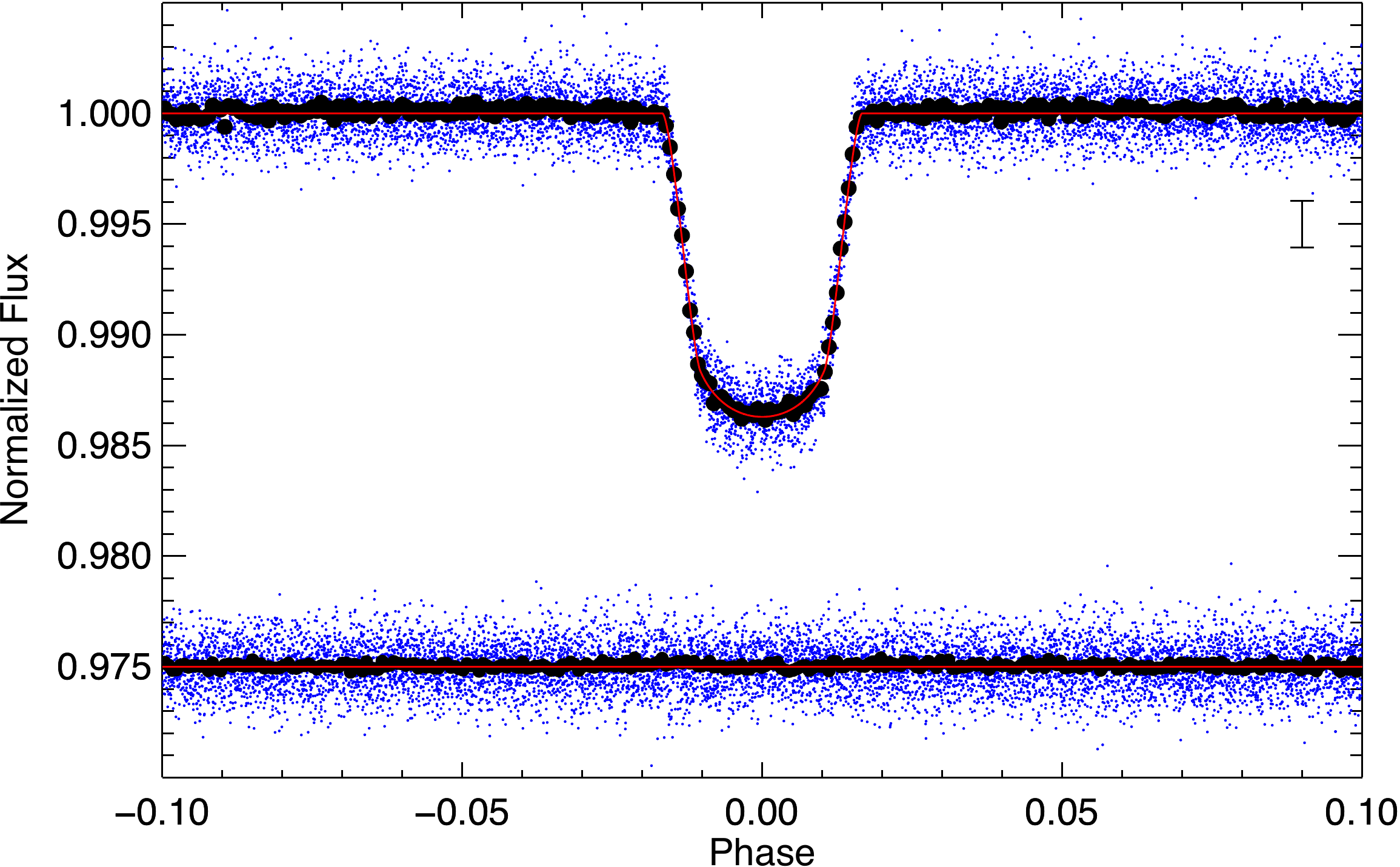}
\caption{Phased folded light curve of XO-6b from TESS. The unbinnned and binned data are shown in blue and black, respectively. The best-fitting model obtained from the EXOplanet MOdeling Package (\texttt{EXOMOP}) is shown as a solid red line. The residuals (light curve - model) are shown below the light curve.} 
\label{fig:modelfit}
\end{figure*}

To obtain the final fitted parameters, the light curve of XO-6b was phase-folded at each individual derived mid-transit time and modeled with \texttt{EXOMOP}. The phase-folded light curve and model fit can be found in Figure \ref{fig:modelfit}. We use the light curve model results combined with literature values to calculate the planetary radius (R$_{b}$), mass (M$_{b}$;  \citealt{Winn2010b}), density ($\rho_{b}$), surface gravity ($\log{g_{b}}$; \citealt{Southworth2007a}), equilibrium temperature (T$_{eq}$), Safronov number ($\Theta$; \citealt{Safronov1972}; \citealt{Southworth2010a}), orbital distance ($a$), inclination, and stellar density ($\rho_{\ast a}$; \citealt{Seager2003}). The period ($P_{orb}$) and transit ephemeris (T$_{C}$[0]) are derived using the mid-transit times from each TESS transit event (Table \ref{tb:lighcurve_model}). Our derived planet properties and transit ephemeris for XO-6b are shown in Table \ref{tb:planet_parameters}. All the planetary parameters are consistent with their discovery values \citep{Crouzet2017} but their precision is greatly improved.        

\begin{table}
\caption{Physical properties of XO-6b derived from the light curve modeling of the TESS data}
    \centering
    \begin{tabular}{lcccc}
     \hline
     Parameter              &  units    & value  & 1 $\sigma$ uncertainty  \\
     \hline
  R$_p$/R$_\ast$            &               & 0.11494&0.00029   \\
  a/R$_\ast$                &               & 8.383&0.074       \\
  Inclination               & \degree       & 85.235&0.087      \\
  Duration         & mins          & 179.94  & 0.11 \\
 b                        &               & 0.696  & 0.014 \\
  R$_{b}$                   & R$_{Jup}$     & 2.08&0.18          \\
  M$_{b}$                   & M$_{Jup}$     & 2.01&0.71                      \\
  $\rho_{b}$                & g cm$^{-3}$  & 0.28&0.12                      \\
  $\log{g_{b}}$   	        &   cgs         & 3.31&0.19                      \\
  $\rho_{\ast a}$           & g cm$^{-3}$   &  0.786 & 0.09642                                 \\
  T$_{eq}$                  &   K           & 1641 & 24                      \\
  $\Theta$                  &               & 0.093& 0.035                      \\ 
  a                         &  au           & 0.0725&0.0063                          \\
  Period    	            &  day          & 3.7649893&0.0000037      \\
  $T_{c}(0)$                & BJD$_{TDB}$   & 2456652.7157 & 0.0022         \\
   \hline
    \end{tabular}
    \label{tb:planet_parameters}
\end{table}

Transit timing variations are conveniently studied in terms of O$-$C, where O is the observed transit time and C is the corresponding calculated transit time. We calculated C with the linear ephemeris formula 

\begin{equation}
T_E = T_0 + P_{orb} \times E     
\end{equation}

\noindent where $T_0$ is the reference transit time, $P_{orb}$ is the orbital period, $E$ is the transit epoch, and $T_E$ is the calculated transit time at epoch $E$. 

For research requiring accurate timing, it is important to consider the clock accuracy and the time standard that are used because ambiguity between different standards can produce spurious timing effects that could be mistaken for TTVs or bias eccentricity measurements \citep{Eastman2010}.  The time stamps of the data products produced by the TESS SPOC pipeline are TESS Julian Dates (TESS JD = JD$-$2457000.0) in the Barycentric Dynamical Time standard, BJD$_{TDB}$\footnote{\url{https://archive.stsci.edu/files/live/sites/mast/files/home/missions-and-data/active-missions/tess/_documents/TESS_Instrument_Handbook_v0.1.pdf}}, which is usually the most accurate time standard to use as it accounts for many different timing corrections, including leap seconds \citep[e.g.,][]{Eastman2010}.  This is achieved with a series of time conversions outlined as follows. Time stamps are first recorded by the 1 Hz spacecraft clock aboard TESS.  While the frequency of this clock drifts due to thermal and aging effects, it is correlated with Coordinated Universal Time (UTC) by the Mission Operations Center at every contact, which occurs approximately every three days. The Payload Operations Center then converts the spacecraft clock counts into TESS Julian Day before passing it to the SPOC for a final conversion into BJD$_{TDB}$\footnote{\url{https://archive.stsci.edu/files/live/sites/mast/files/home/missions-and-data/active-missions/tess/_documents/TESS_Instrument_Handbook_v0.1.pdf}}.  This final conversion is implemented with the same algorithm that was used for \textit{Kepler}, and works by using the Navigation and Ancillary Interface (NAIF) SPICE kernel trajectory file\footnote{\url{https://naif.jpl.nasa.gov/naif/index.html}} to calculate the projected distance to the solar system barycenter and the resulting timing corrections for each star in the TESS field of view (Jenkins, J., private communication).  By analysing contemporaneous TESS and ground-based observations of several binary star systems, \cite{vonEssen2020} showed that the TESS BJD$_{TDB}$ times are correct to an absolute precision of $<$6 seconds.

\begin{figure*}
\centering
  \sidesubfloat[\textbf{(a.)}]{\includegraphics[width=0.65\textwidth,page=1]{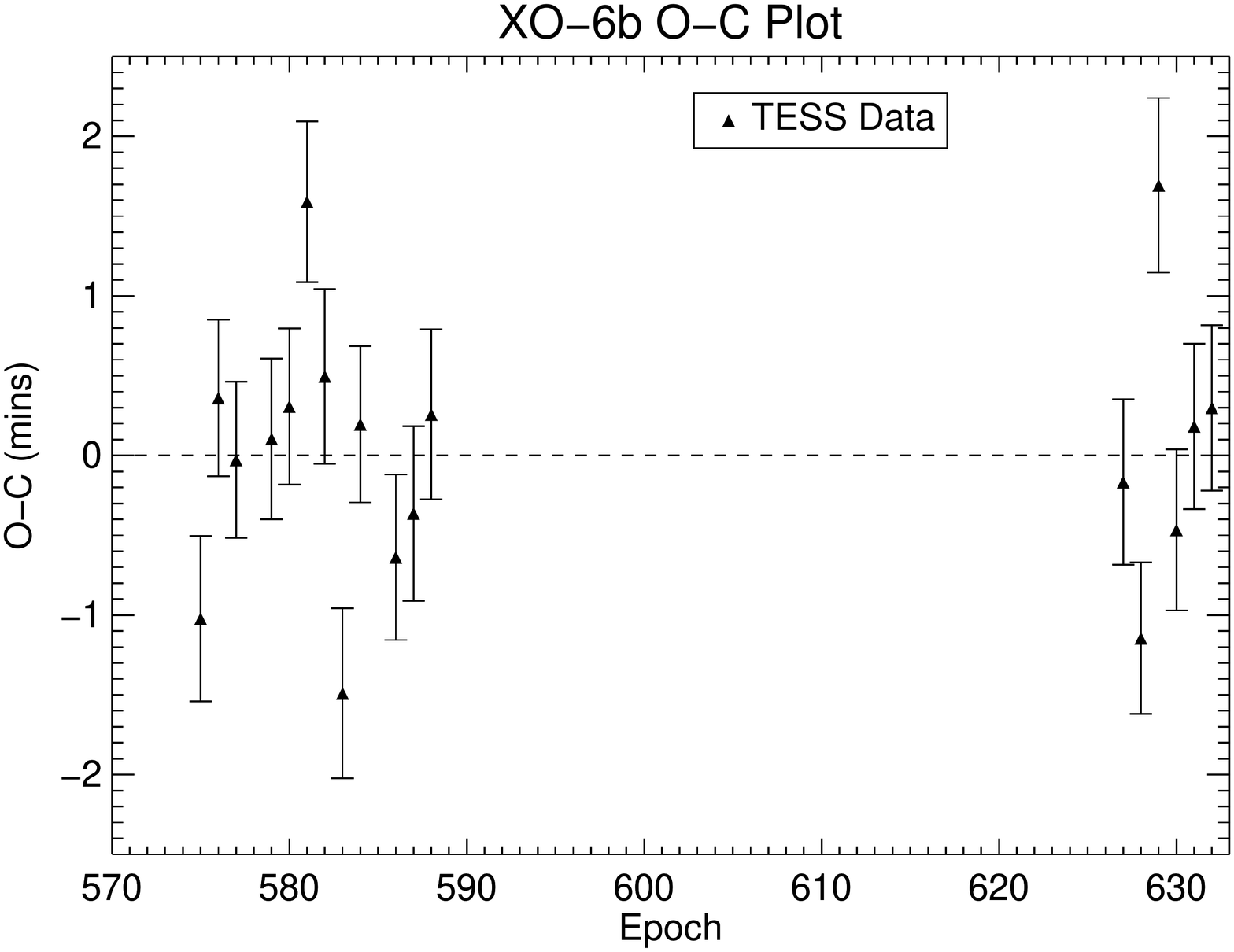}} \\
    \sidesubfloat[\textbf{(b.)}]{\includegraphics[width=0.65\textwidth,page=1]{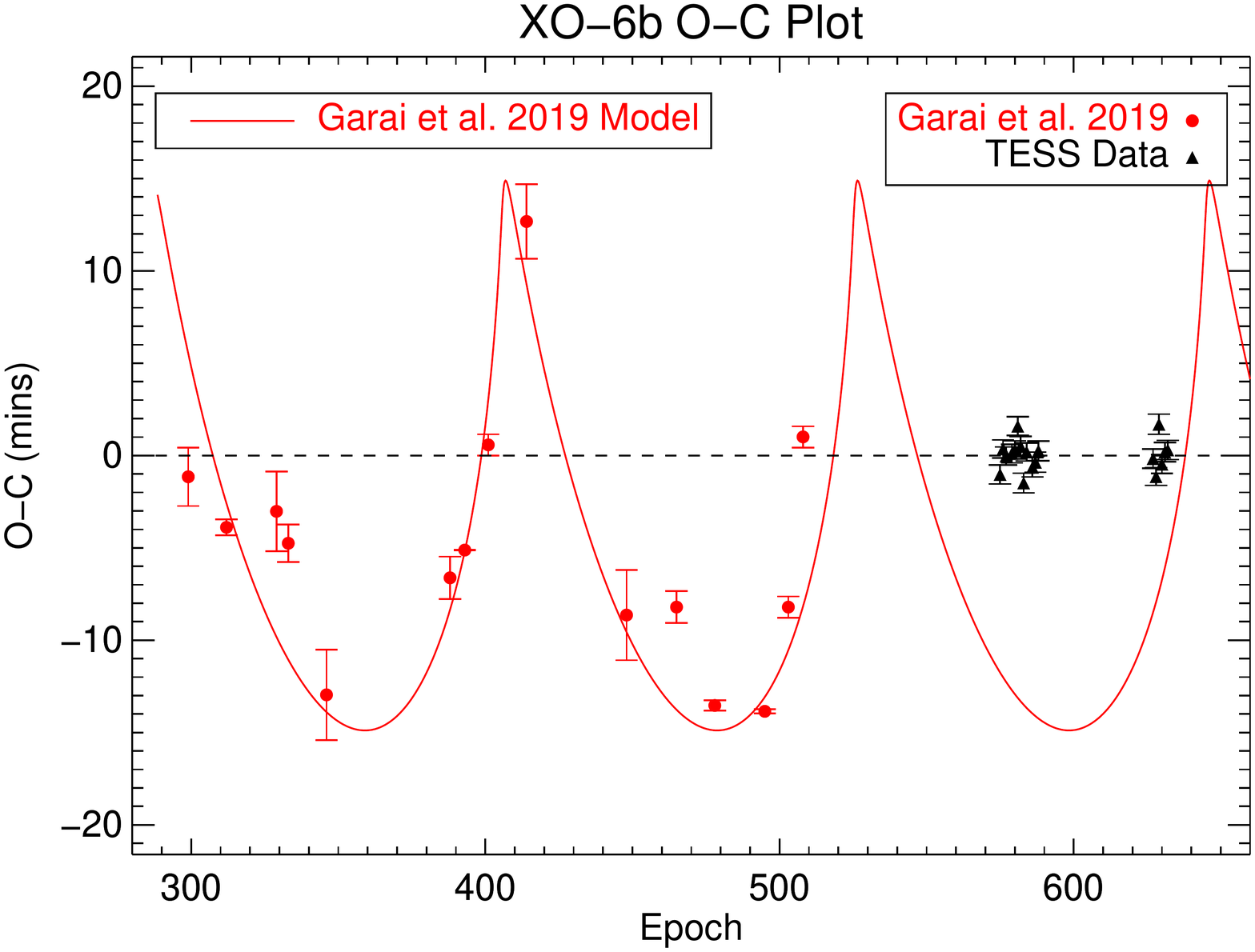}} 
\caption{a.) The observation minus calculation mid-transit time (O-C) diagram derived using the TESS data (black). b.) The O-C diagram including data points from \citet{Garai2020} (red). The best-fit light-time effect (LiTE) model from \citet{Garai2020} is shown as a red line.}
\label{fig:OC}
\end{figure*}

Figure \ref{fig:OC} shows the O$-$C values we derived from the TESS observations, along with the O$-$C values and best fit light-time effect (LiTE) model 
from \cite{Garai2020}. To compare our transit times to this model, we used the publicly available \texttt{OCFit} package\footnote{\url{https://github.com/pavolgaj/OCFit}} \citep{Gajdos2019OCFitRef} to reproduce it using their fitted parameters shown in Table \ref{tb:Garai_LiTE_model_params} in Appendix \ref{app:Garai2020_model}.  While this model is a good fit to their reported TTVs, they did not observe radial velocities consistent with the XO-6 star-planet system having a stellar mass companion capable of inducing these light-time effects.  Therefore, they favor the interpretation that these TTVs are caused by resonant perturbations between XO-6b and another unknown low-mass planet in the system, although they do not quantify the mass of their suggested perturbing planet.

\section{Results and discussion} \label{sec:results}

Our analysis of the TESS data of XO-6b yielded an updated orbital period and a transit epoch of \NewPeriod{} and \NewTransitEpoch{}, respectively.

Notably, we did not find any evidence of transit timing variations in the case of XO-6b, despite the TESS data having sufficient precision and time baseline to clearly reveal the TTVs reported by \citet{Garai2020}.  Specifically, the 3$\sigma$ upper-limit on TTVs allowed by the TESS observations is \TessTTVlim{}, while their LiTE model predicts TTVs of 12-14 minutes in Sectors 19 and 20 and 5-9 minutes in Sector 26. Therefore, the TESS data rule out their LiTE model by 14-16$\sigma$ in Sectors 19 and 20 and 6-11$\sigma$ in Sector 26. While the cause of this discrepancy is not clear, here we discuss some possible explanations.

The analysis of \cite{Garai2020} utilized 15 transits, of which 8 were observed by the authors themselves with telescopes at the Astronomical Institute of the Slovak Academy of Sciences and 7 were adopted from the ETD after having been observed by others. While the data obtained from the ETD do not exhibit stronger TTVs than their own observations (see Figure \ref{fig:barycor} in Appendix \ref{app:barycor_plot}) and they only used data from the ETD that had clearly indicated timing epochs (either JD or HJD) and good quality, it is still possible these data points introduced unaccounted for uncertainties that contributed to the discrepancy. We used the \texttt{astropy.time}\footnote{\url{https://docs.astropy.org/en/stable/time/}} package to calculate the barycentric correction for the data used by \citet{Garai2020} and found that it ranges from $-$5 to $+$5 minutes, with a mean value of $+$0.8 minutes (see Figure \ref{fig:barycor} in Appendix \ref{app:barycor_plot}). Therefore, while a few of the smaller TTVs they reported could be related to barycentric corrections, the larger ones must have other causes.

\cite{Garai2020} also utilized spectroscopic radial velocity (RV) observations, which showed no evidence of an additional body in the XO-6 system. Furthermore, simultaneous fits to their RV and O-C data did not produce a consistent solution, potentially suggesting spurious effects.

\citet{Crouzet2017} found no evidence of XO-6b exhibiting TTVs, which \citet{Garai2020} attributed to the \citet{Crouzet2017} observations covering a relatively short time span, having non-optimal time sampling, and/or higher uncertainties in their O-C values.  However, the results of \citet{Crouzet2017} and \citet{Garai2020} could also be reconciled if the findings of \citet{Garai2020} were affected by unknown timing errors. 

The presence or absence of companion planets near a hot Jupiter are indicative of how it formed \citep{Steffen2012}.  The vast majority of transiting hot Jupiters are the only transiting planet in their system \citep[e.g.,][]{Huang2016} so our interpretation that XO-6b does not have a companion is consistent with this clear trend. The trend could arise either because the systems do not contain any other planets or because they contain planets on inclined orbits that do not transit. 

\citet{Spalding2016} showed that hot stars like XO-6 with T$_{eff}$ $>$ 6200 K can exert torques on their planets that increase their inclinations so that they do not transit. This effect is stronger for closer-in planets, potentially resulting in a planetary system harboring only one transiting planet.  

\citet{Batygin2016} argue that a substantial fraction of the hot-Jupiter population formed in situ.  They showed that this process would lead frequently to hot-Jupiters being accompanied by low-mass planets with periods shorter than approximately 100 days.  Furthermore, they found that that dynamical interactions early in the systems' lifetimes should increase the inclinations of these companions, making them unlikely to transit.

While the presence of an additional non-transiting planet in the XO-6 system would be consistent with the interpretation of \citet{Garai2020}, \cite{Nesvorny2009} showed that the TTV signal caused by inclined non-transiting companions may be more pronounced than that caused by co-planar companions.  Therefore, if XO-6b does have a non-coplanar, non-transiting companion, it is reasonable to expect that it could be detected.  Rigorous modelling could constrain the allowed parameter space of a potential candidate while still being consistent with XO-6b not showing any transit timing variations \citep[e.g.,][]{Hrudkova2010}, but that is beyond the scope of the present study.

Alternatively, \citet{Mustill2015} show that if a hot Jupiter reaches its current orbit by high-eccentricity migration, any inner low-mass planets will collide with their host star or the migrating giant planet and be destroyed.  The initial high eccentricity required for this process can be produced by planet-planet scattering, the Kozai effect or low-inclination secular interactions.  Once the giant planet's pericenter has migrated to within a few hundredths of an AU from its host star, its orbit will be tidally circularized, resulting in a hot Jupiter with a very low eccentricity, such as XO-6b.    

Given the lack of evidence for an additional non-transiting planet in the system, we consider it likely that XO-6b migrated inwards to its current orbital distance. Future observational constraints on XO-6b's atmospheric C/O ratio, which would be indicative of where a planet formed in a protoplanetary disk, could confirm such a scenario  \citep[e.g.,][]{Oberg2011,Cridland2019}.

\section{Conclusion} \label{sec:conclusion}

We utilized publicly available TESS data to further characterize the hot-Jupiter XO-6b.  Our analysis yielded an updated period of \NewPeriod{} and transit epoch of \NewTransitEpoch{}. Moreover, we found no evidence of the transit timing variations reported by \cite{Garai2020}, despite the precision and time baseline of the TESS data being sufficient to reveal them, highlighting the usefulness of TESS follow-up observations of interesting targets found with ground-based observations. The cause of the tension between our results and those of \citet{Garai2020} is not clear but it may be due to unknown timing errors in their ground-based data. This underscores the importance of careful absolute telescope clock calibrations and considerations of timing standards \citep{Eastman2010}, which is not only necessary for TTV analysis, but also for constraining ephemerides well enough to enable efficient scheduling of atmospheric characterization observations on high-demand telescopes like the upcoming James Webb Space Telescope and the 30-meter class telescopes \citep{Dragomir2020,Zellem2020}.

\acknowledgments
We thank Thomas Barclay and Jon Jenkins for assisting us to fully utilize the TESS/SPOC data products, and Ernst de Mooij for helpful discussions.We also thank the anonymous referee for their constructive suggestions.
This paper includes data collected by the TESS mission, which are publicly available from the Mikulski Archive for Space Telescopes  (MAST). Funding for the TESS mission is provided by the NASA Explorer Program.

This research has made use of the Extrasolar Planet Encyclopaedia, NASA's Astrophysics Data System Bibliographic Services, and the NASA Exoplanet Archive, which is operated by the California Institute of Technology, under contract with the National Aeronautics and Space Administration under the Exoplanet Exploration Program.

%

\vspace{5mm}
\facilities{TESS}


\software{\texttt{astropy} \citep{2013A&A...558A..33A},  
          \texttt{EXOMOP} \citep{Pearson2014,Turner2016,Turner2017}, \texttt{OCFit} \citep{Gajdos2019OCFitRef}, \texttt{exoplanet} package \citep{exoplanet}, \texttt{Everest} package \citep{Luger2016}
          }



\newpage 
\clearpage
\appendix

\section{Difference in mid-transit times between different detrending models} \label{app:diff_detrend}

\begin{figure}[htb]
\plotone{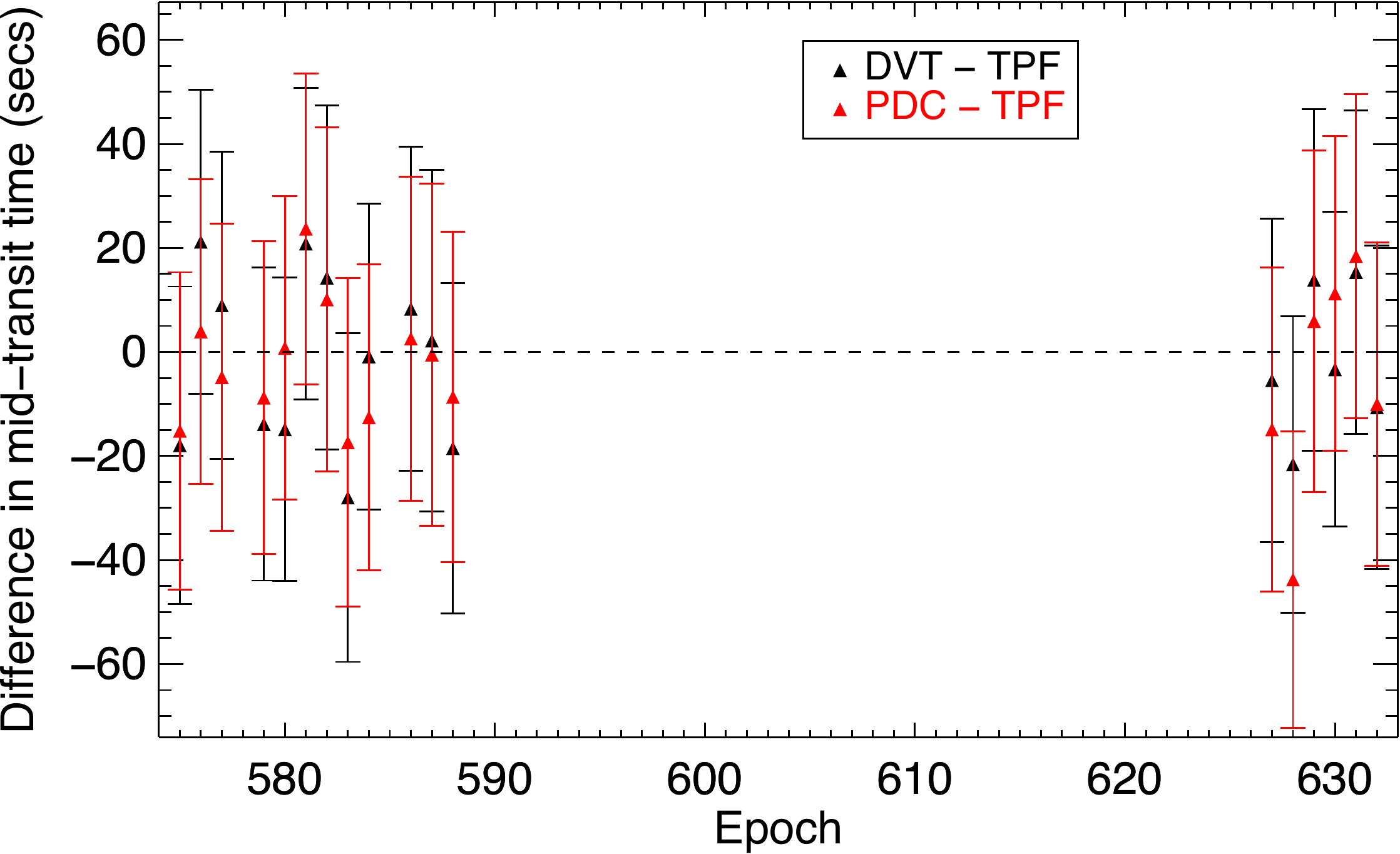}
\caption{Differences in derived mid-transit times relative to our detrended light curve produced from the target pixel files (TPFs) for the data validation timeseries (DVT) (black) and the Presearch Data Conditioning (PDCSAP\_FLUX) (red). 
\label{fig:LC_comp}}
\end{figure}

\newpage 
\clearpage
\section{Transit fits to individual TESS transit events} \label{app:individual_transits}

\begin{table}[htb]
\caption{Individual TESS transit parameters for XO-6b derived using \texttt{EXOMOP} }
    \centering
    \begin{tabular}{lccc}
    \hline
    \hline 
    \multicolumn{4}{c}{Sector 19} \\
    \hline 
    \hline 
    Transit                 & 1                     & 2                     & 3                                \\
    T$_{c}$ (BJD$_{TDB}$-2458810)   & 7.58383$\pm$0.00036   &11.34978$\pm$0.00033   & 15.11450$\pm$0.00030    \\
    R$_p$/R$_\ast$          & 0.1141$\pm$0.0011     &0.1148$\pm$0.0011      & 0.1130$\pm$0.0010               \\
    a/R$_\ast$              & 8.73$\pm$0.36         &8.49$\pm$0.33          & 8.91$\pm$0.27                  \\
    Inclination (\degree)   & 85.62$\pm$0.39        &85.25$\pm$0.37         &85.83$\pm$0.32         \\
    Duration (days)         & 174.02$\pm$3.05       &174.02$\pm$2.88        &174.02$\pm$2.83        \\
    \hline
    Transit                 &4                     & 5                     & 6   \\
    T$_{c}$ (BJD$_{TDB}$-2458810)   &22.64457$\pm$0.00035  &26.40970$\pm$0.00032   &30.17558$\pm$0.00033  \\
    R$_p$/R$_\ast$          &0.1158$\pm$0.0011     & 0.1159$\pm$0.0014     & 0.1133$\pm$0.0012  \\ 
    a/R$_\ast$              &8.38$\pm$0.29         & 8.64$\pm$0.37         & 8.62$\pm$0.34   \\
    Inclination (\degree)   &85.21$\pm$0.34        & 85.44$\pm$0.43        &85.55$\pm$0.41           \\
    Duration (days)         &177.89$\pm$2.83       & 174.02$\pm$2.87       & 175.96$\pm$2.90              \\ 
    
    \hline 
    \hline  
    \multicolumn{4}{c}{Sector 20} \\
    \hline 
    \hline 
     Transit                 & 1                    & 2                     & 3   \\
    T$_{c}$ (BJD$_{TDB}$-2458810)   & 33.93981$\pm$0.00035  & 37.70342$\pm$0.00030  & 41.46958$\pm$0.00032                         \\
    R$_p$/R$_\ast$          & 0.1154$\pm$0.0012     & 0.11425$\pm$0.00088   & 0.11550$\pm$0.00082       \\
    a/R$_\ast$              & 7.95$\pm$0.27         & 8.59$\pm$0.249        & 8.13$\pm$     \\
    Inclination (\degree)   & 84.75$\pm$0.35        &85.51$\pm$0.27         &84.96$\pm$0.24\\
    Duration (days)         & 183.87$\pm$2.86       & 175.96$\pm$2.86       &181.93$\pm$2.83\\
   \hline
    Transit                 & 4                     & 5                     & 6    \\
    T$_{c}$ (BJD$_{TDB}$-2458810)   & 48.99898$\pm$0.00034  & 52.76416$\pm$0.00037  & 56.52958$\pm$0.00036                       \\
    R$_p$/R$_\ast$          & 0.1143$\pm$0.0012     &0.11789$\pm$0.000 99   &0.1153$\pm$0.0011 \\
    a/R$_\ast$              & 8.55$\pm$0.36         &7.58$\pm$0.23         &8.09$\pm$0.28                  \\
    Inclination (\degree)   & 85.44$\pm$0.40        &84.23$\pm$0.30         & 84.91$\pm$0.35               \\
    Duration (days)         & 176.13$\pm$2.85       &184.04$\pm$2.87        &180.00$\pm$2.85                \\
    \hline
    \hline 
    \multicolumn{4}{c}{Sector 26} \\
    \hline 
    Transit                 & 1                         & 2                     & 3   \\
    T$_{c}$ (BJD$_{TDB}$-2458810)   & 203.36387$\pm$0.00036     & 207.12818$\pm$0.00033 & 210.89514$\pm$0.00038       \\
    R$_p$/R$_\ast$          & 0.1142$\pm$0.0015         &0.1149$\pm$0.0012      & 0.1149$\pm$0.0012          \\
    a/R$_\ast$              & 8.55$\pm$0.43             &8.52$\pm$0.33          &8.29$\pm$0.35  \\
    Inclination (\degree)   & 85.50$\pm$0.52            &85.41$\pm$0.40         &85.10$\pm$0.41\\
        Duration (days)     & 178.07$\pm$2.88           &177.89$\pm$2.85        &178.07$\pm$2.92\\
    \hline 
    Transit                 & 4                     & 5                     & 6     \\
    T$_{c}$ (BJD$_{TDB}$-2458810)   & 214.65863$\pm$0.00035 & 218.42407$\pm$0.00036 & 222.18914$\pm$0.00036         \\
    R$_p$/R$_\ast$          & 0.1169$\pm$0.0011     & 0.1135$\pm$0.0010     & 0.11576$\pm$0.00097                 \\
    a/R$_\ast$              & 8.47$\pm$0.26         &8.66$\pm$0.26          & 8.13$\pm$0.21   \\
    Inclination (\degree)   & 85.30$\pm$0.31        &85.57$\pm$0.30         & 84.92$\pm$0.25   \\
    Duration (days)         & 175.96$\pm$2.86       &174.02$\pm$2.91        & 180.00$\pm$2.88  \\
    \hline
    \hline 
    \end{tabular}
    \tablecomments{The linear and quadratic limb darkening coefficient used in the analysis are 0.3158 and 0.2206 \citep{Claret2017}}
    \label{tb:lighcurve_model}
\end{table}

\begin{figure*}
\centering
\begin{tabular}{cc}
 \includegraphics[width=0.50\textwidth]{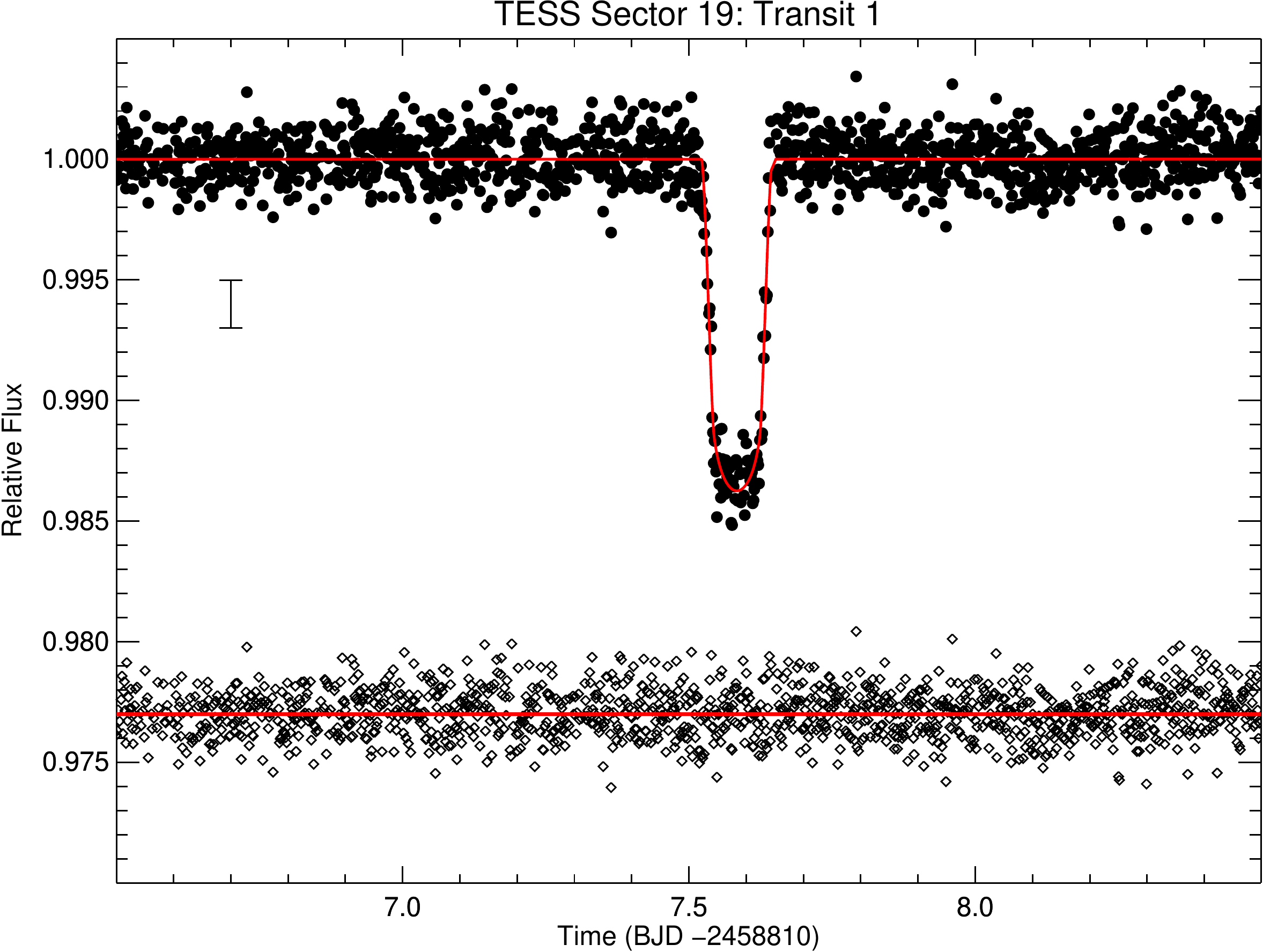} & 
  \includegraphics[width=0.50\textwidth]{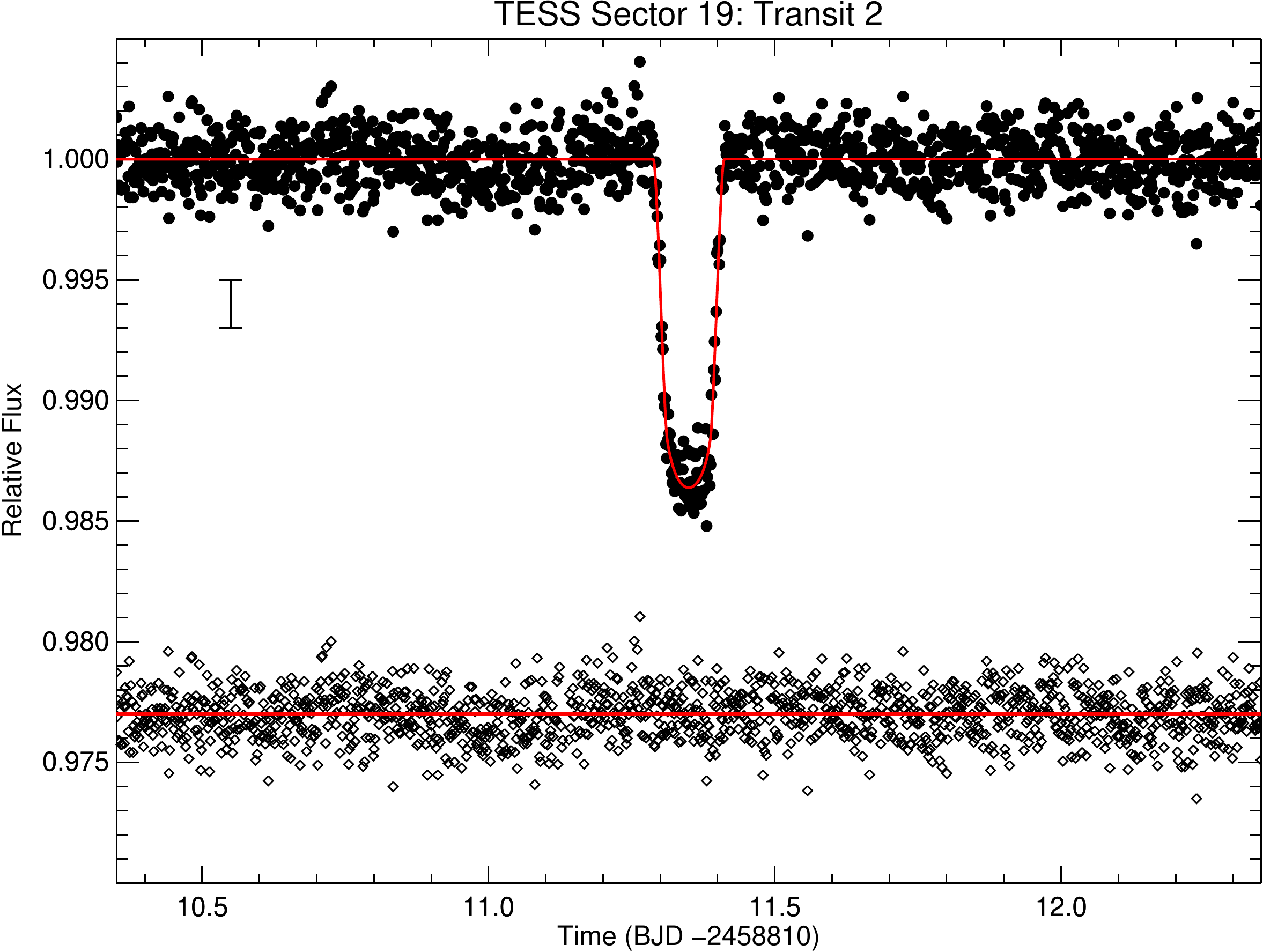} \\
    \includegraphics[width=0.50\textwidth]{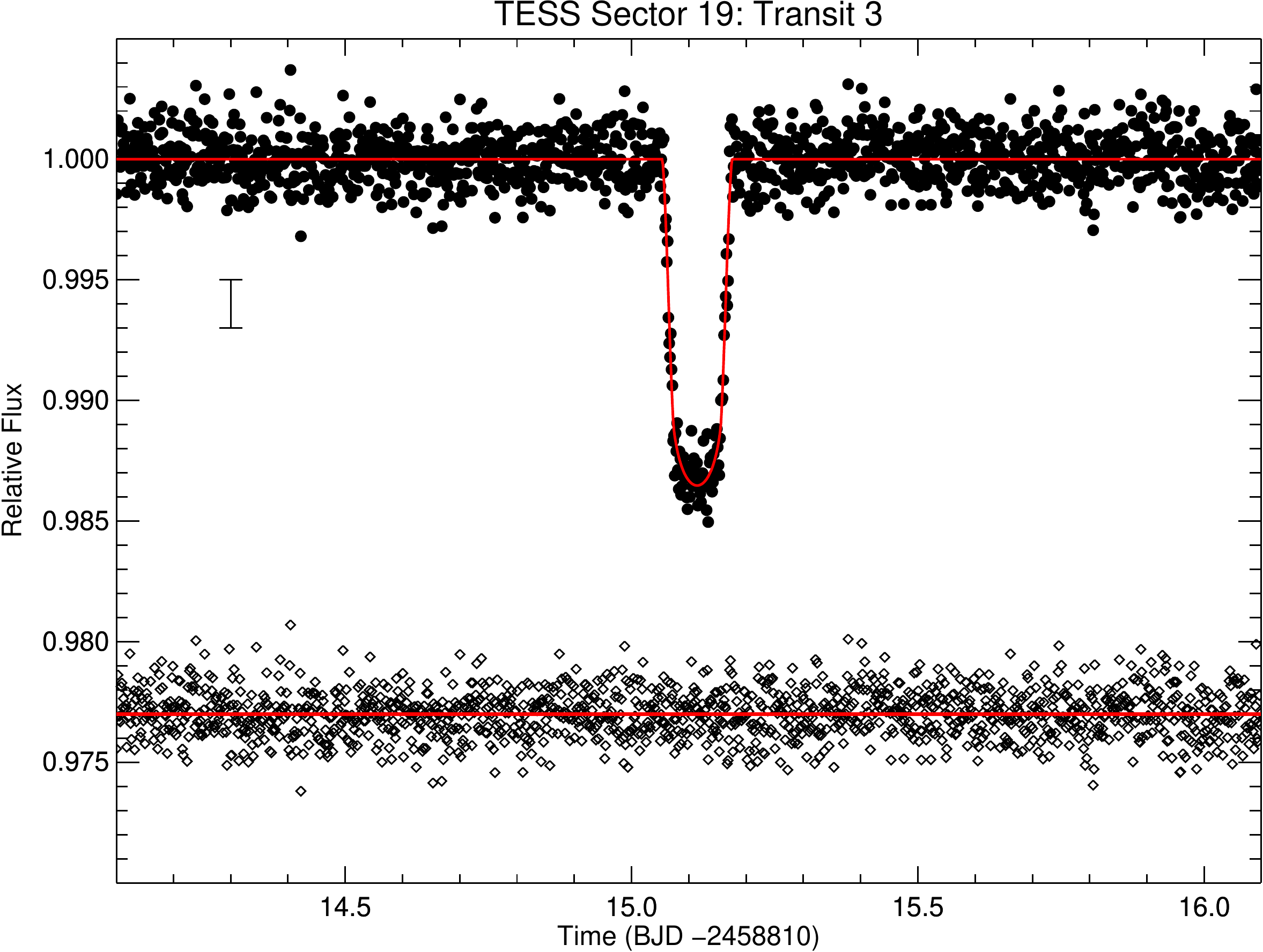} &
 \includegraphics[width=0.50\textwidth]{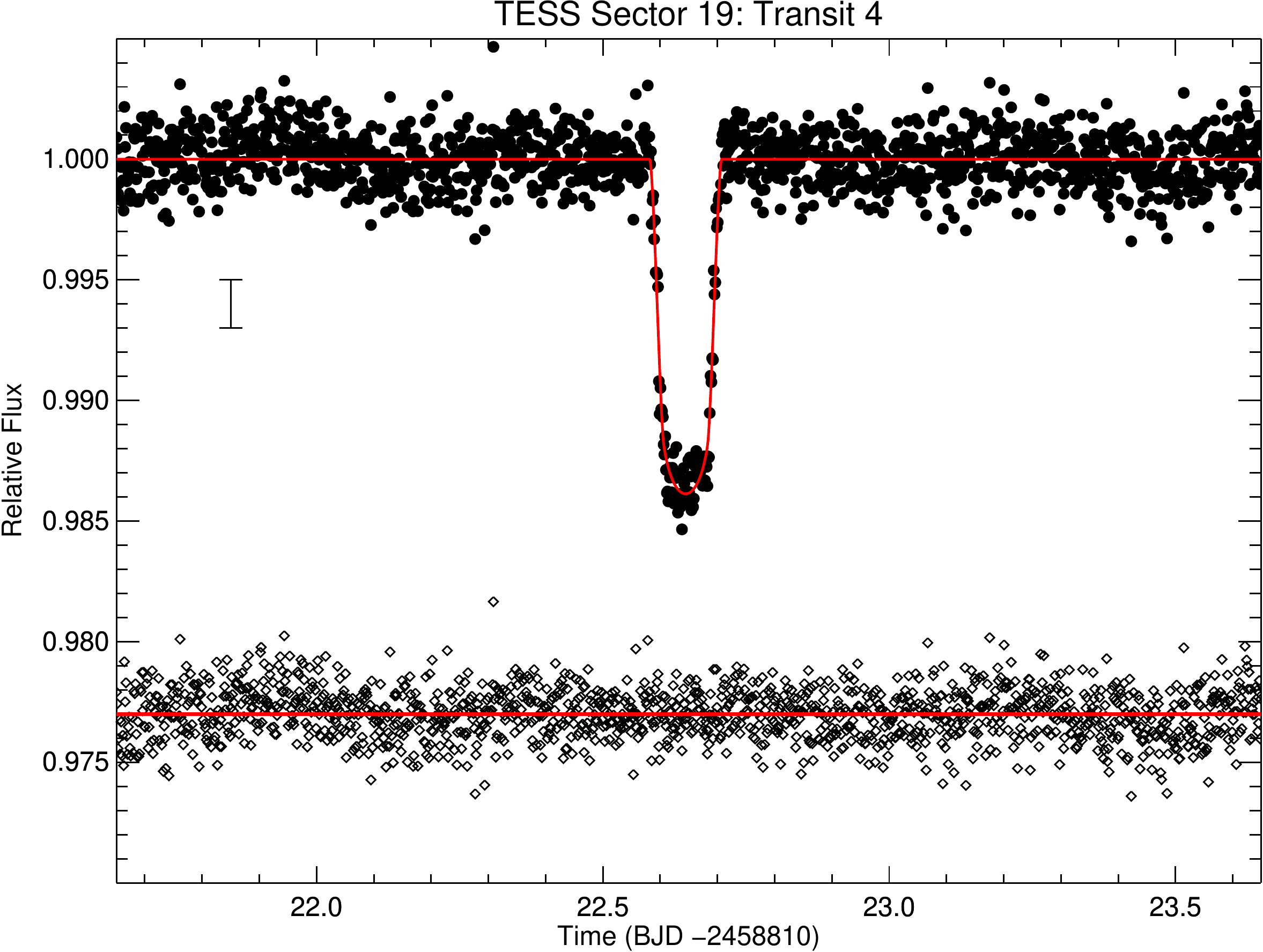} \\
  \includegraphics[width=0.50\textwidth]{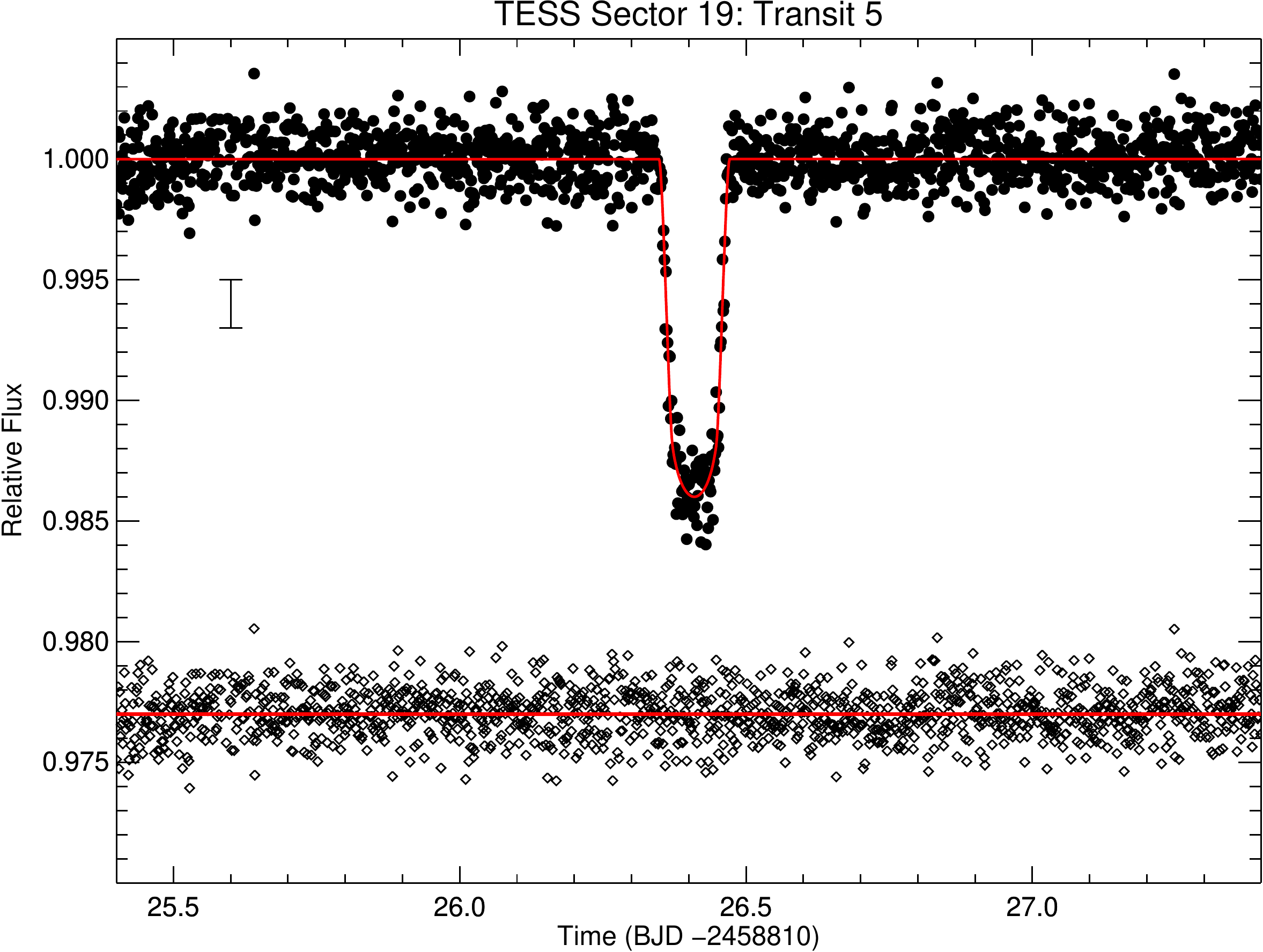} &
    \includegraphics[width=0.50\textwidth]{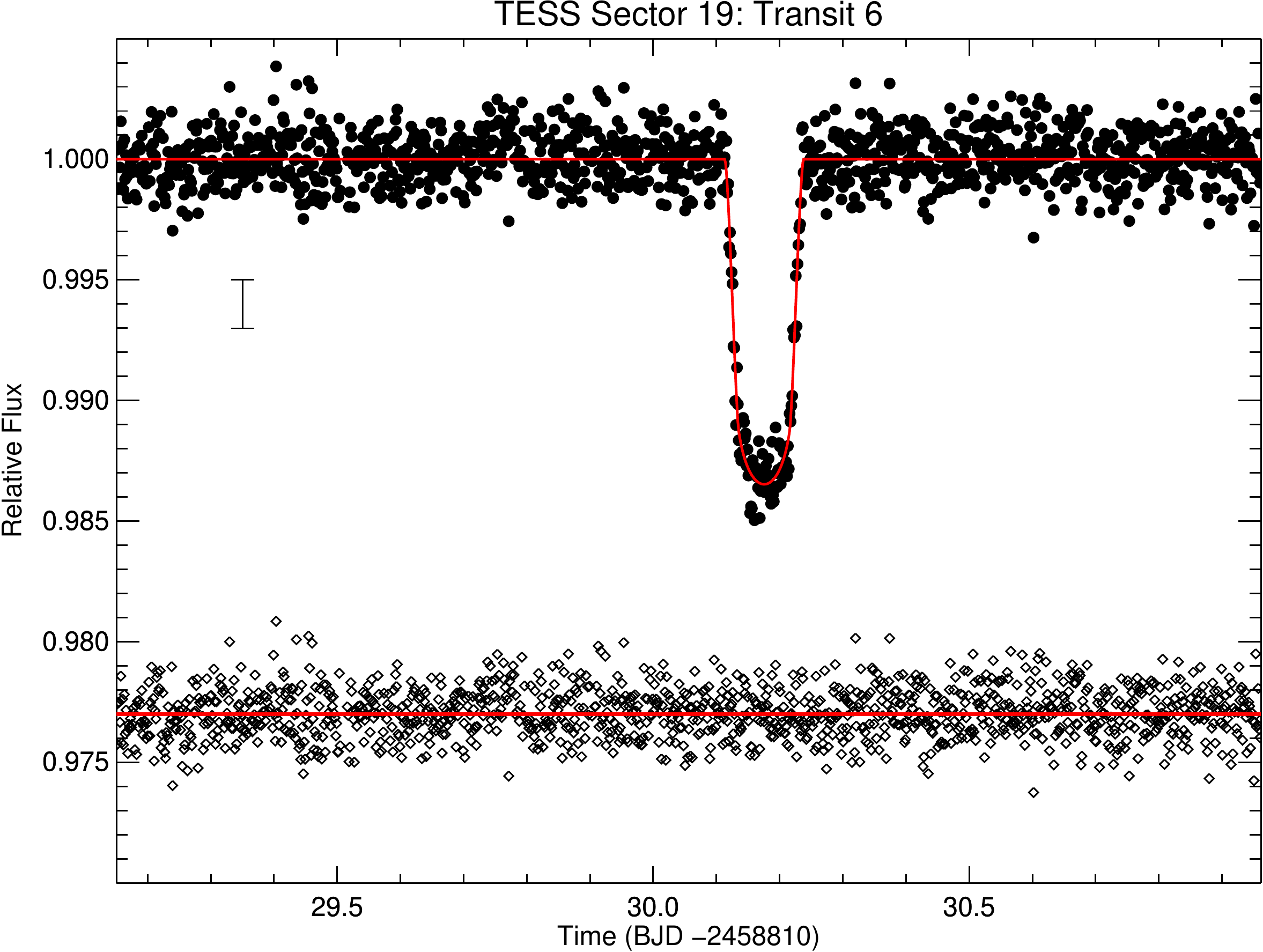} \\
  \end{tabular}
\caption{Individual TESS transit events from Sector 19 of XO-6b. The best-fitting model obtained from the EXOplanet MOdeling Package (\texttt{EXOMOP}) is shown as a solid red line. The residuals (light curve - model) are shown below the light curve.}
\label{fig:ind_transits_Sec19}
\end{figure*}

\begin{figure*}
\centering
\begin{tabular}{cc}
 \includegraphics[width=0.50\textwidth]{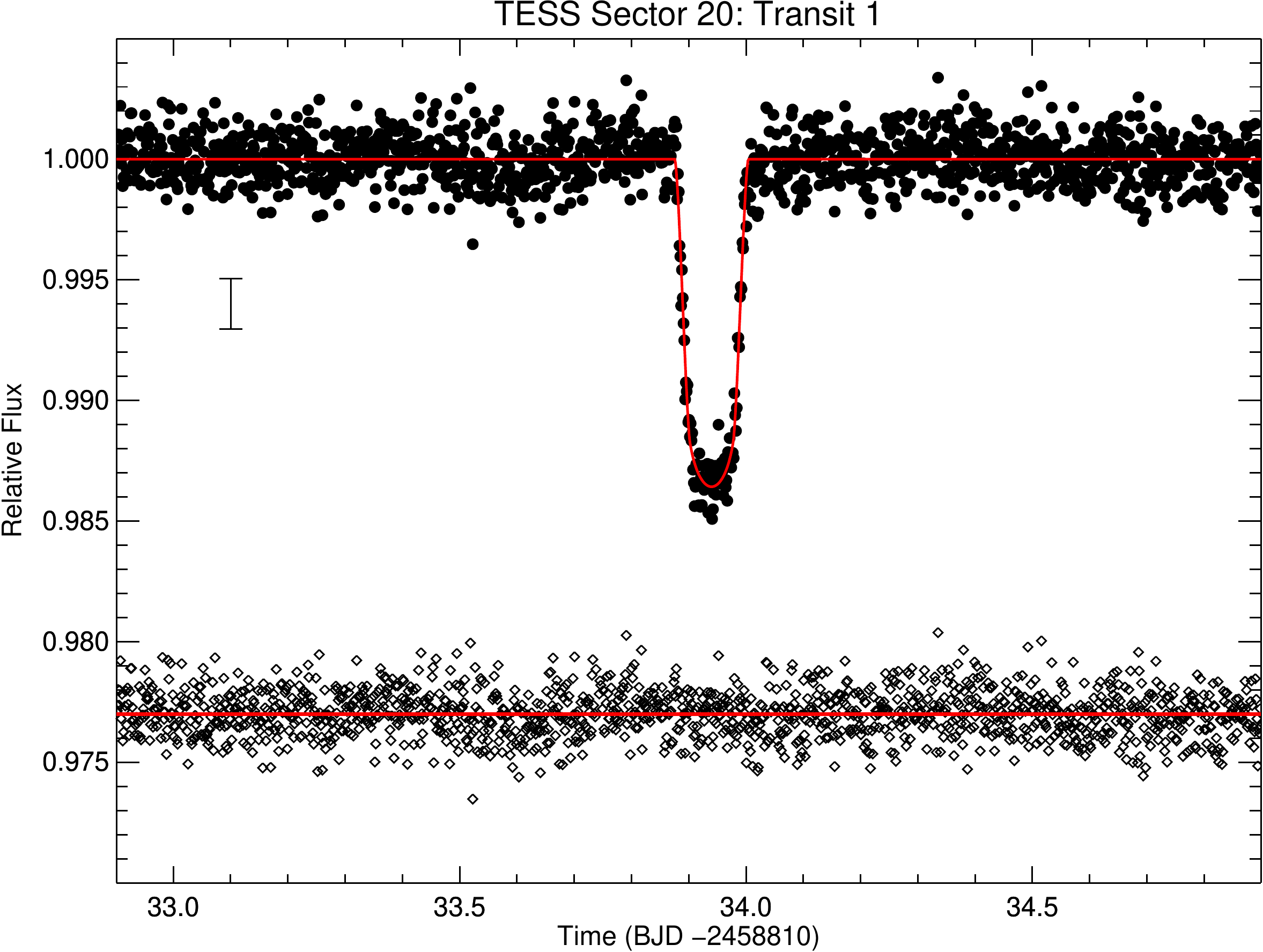} & 
  \includegraphics[width=0.50\textwidth]{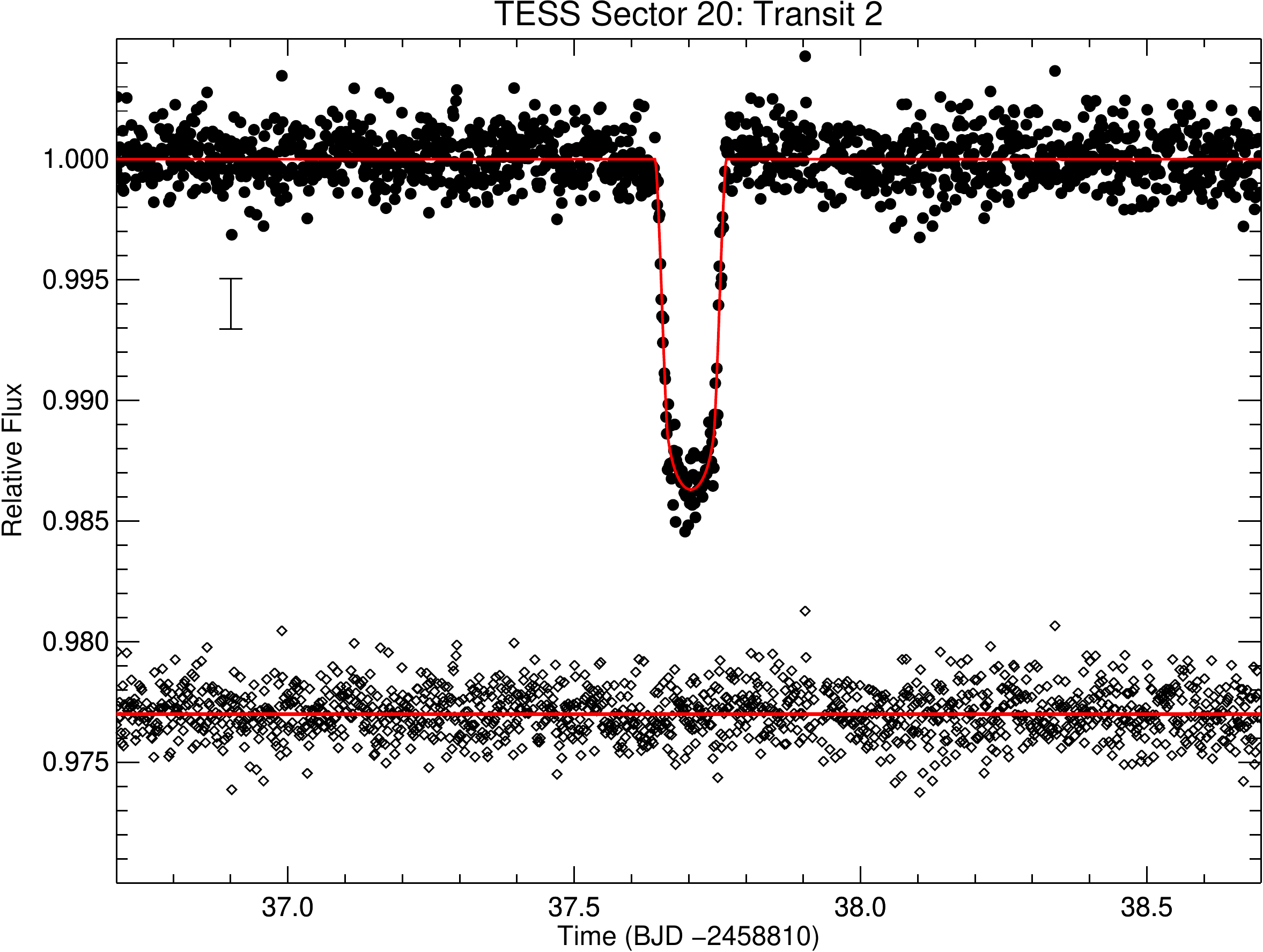} \\
    \includegraphics[width=0.50\textwidth]{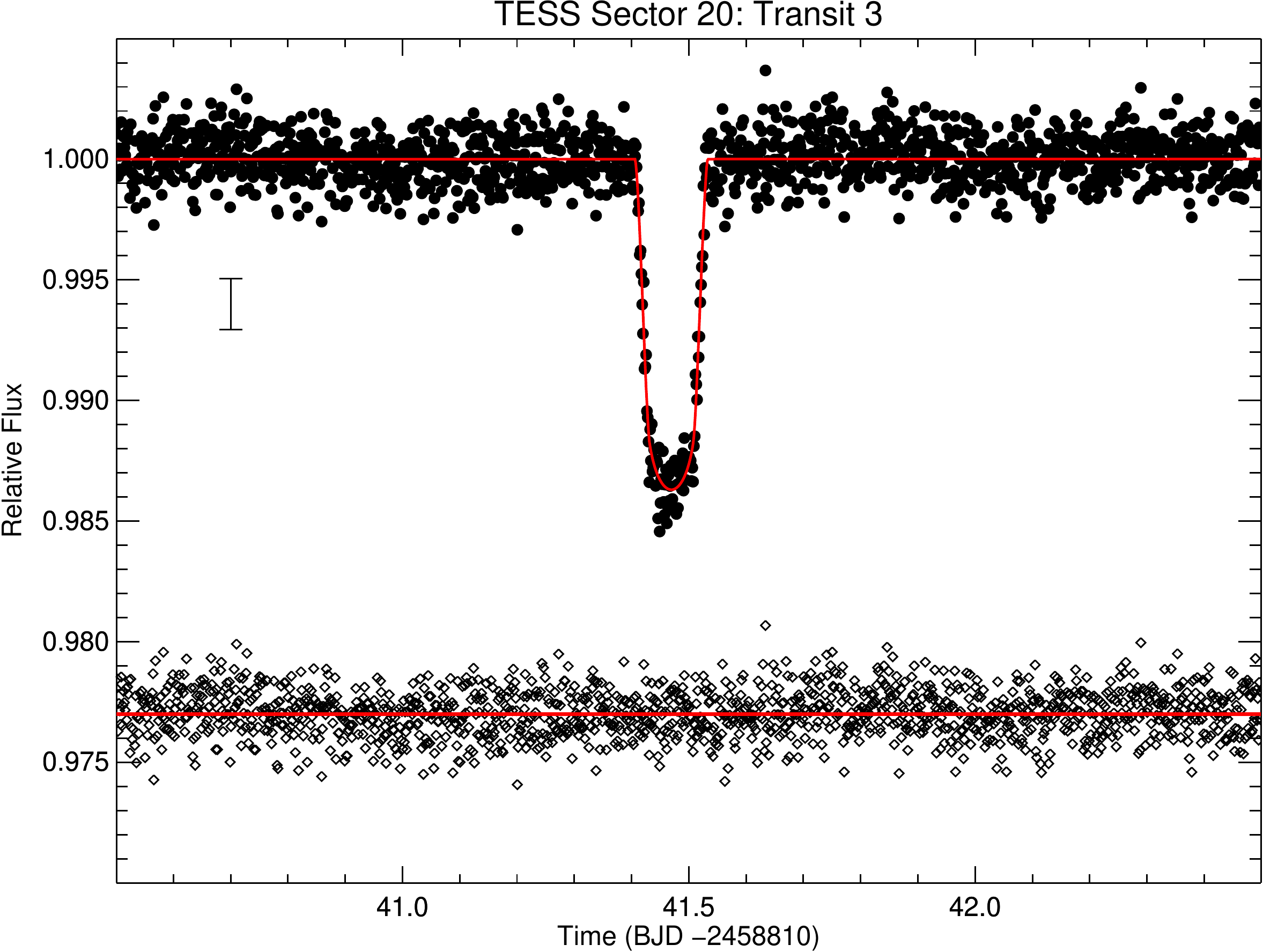} &
 \includegraphics[width=0.50\textwidth]{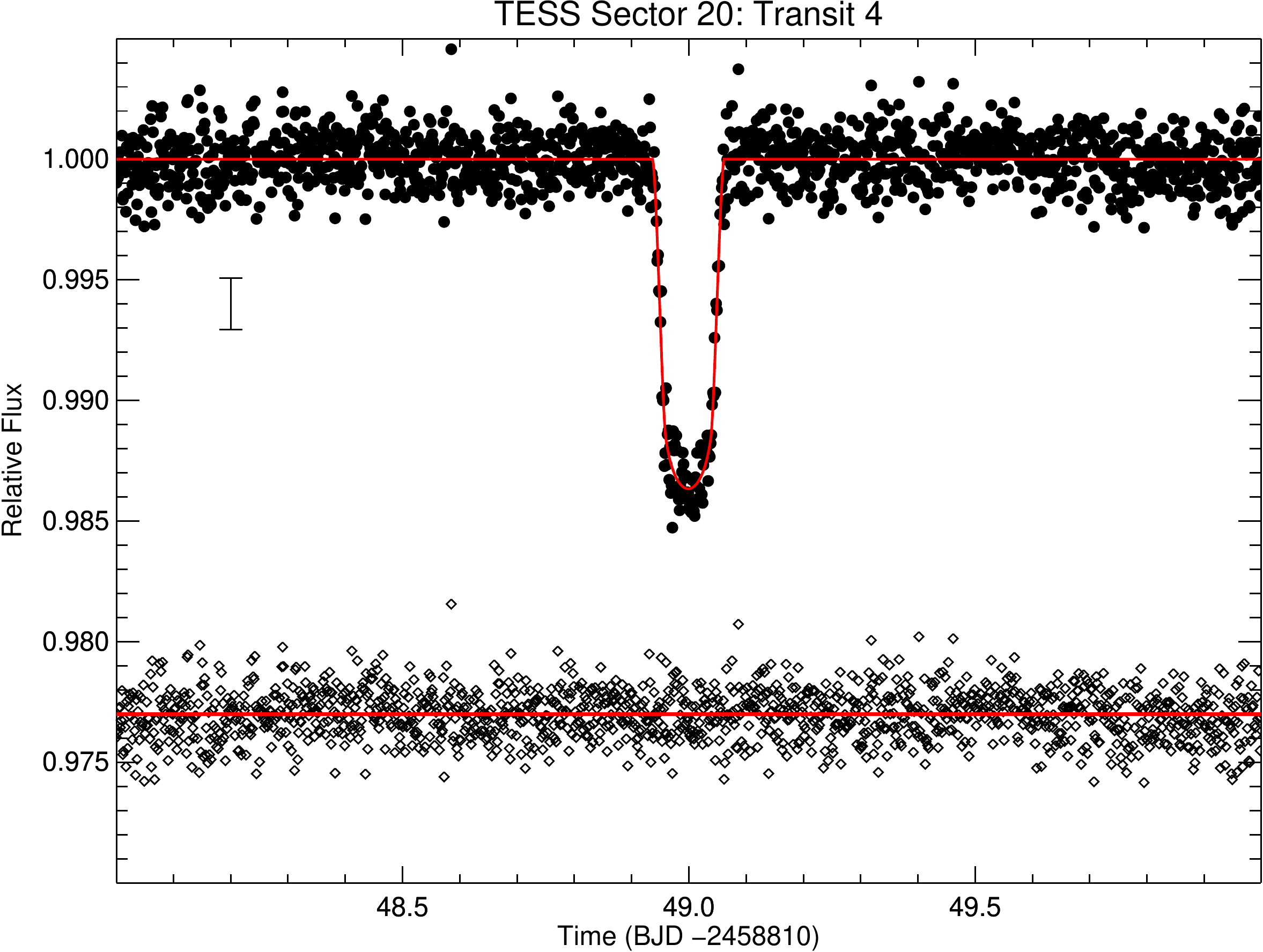} \\
  \includegraphics[width=0.50\textwidth]{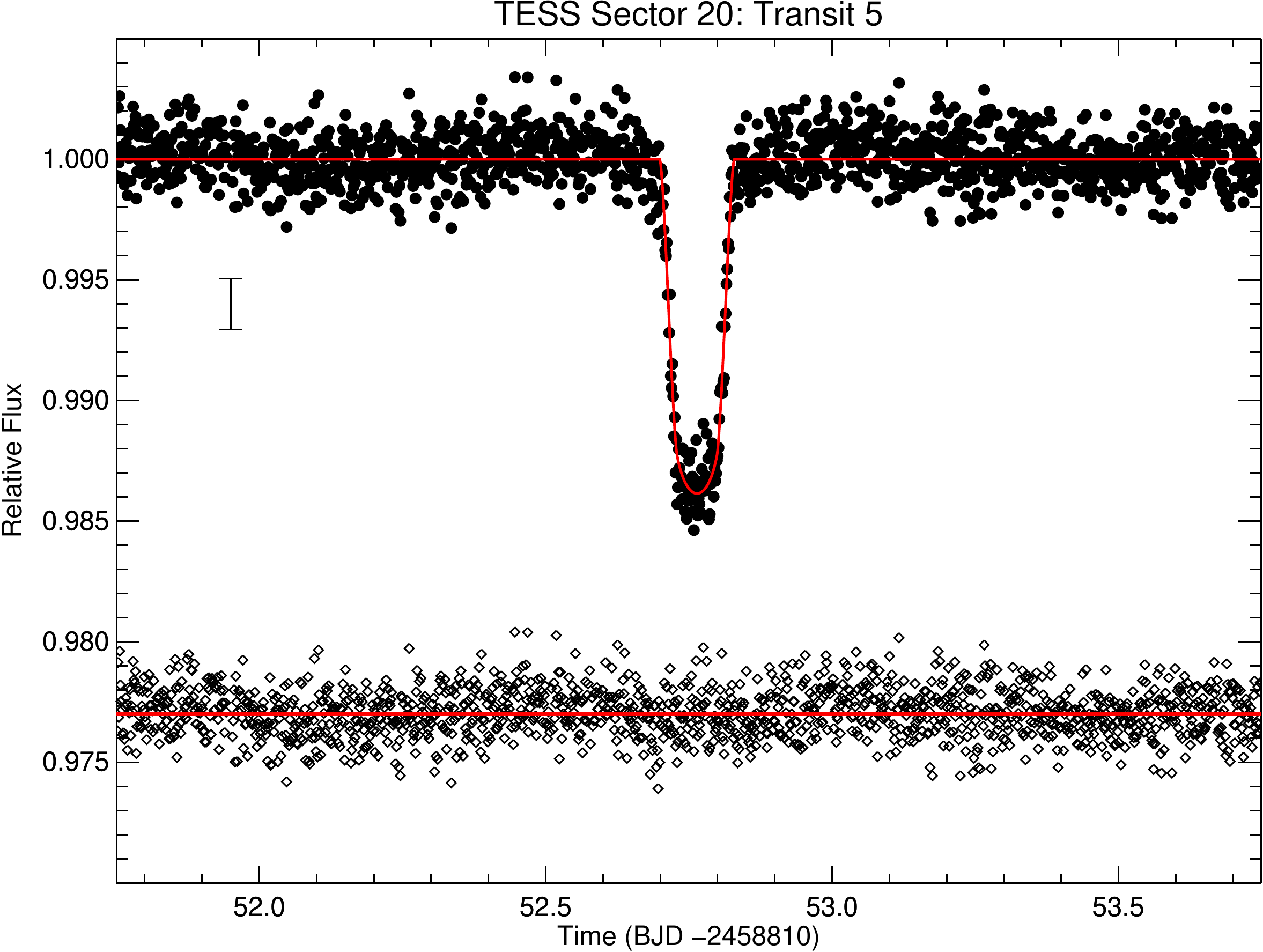} &
    \includegraphics[width=0.50\textwidth]{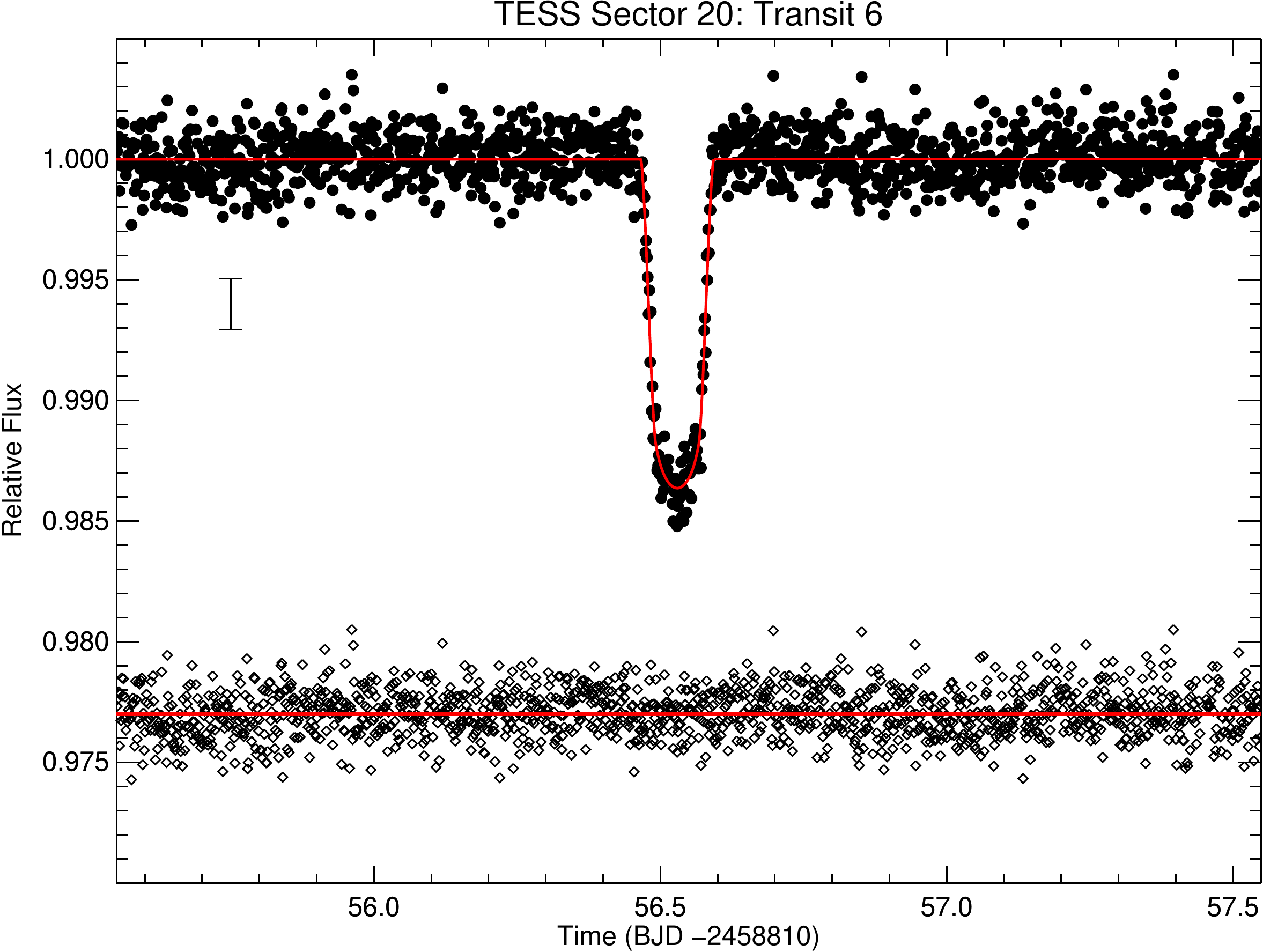} \\
  \end{tabular}
\caption{Individual TESS transit events from Sector 20 of XO-6b. The best-fitting model obtained from the EXOplanet MOdeling Package (\texttt{EXOMOP}) is shown as a solid red line. The residuals (light curve - model) are shown below the light curve.}
\label{fig:ind_transits_sec20}
\end{figure*}

\begin{figure*}
\centering
\begin{tabular}{cc}
 \includegraphics[width=0.50\textwidth]{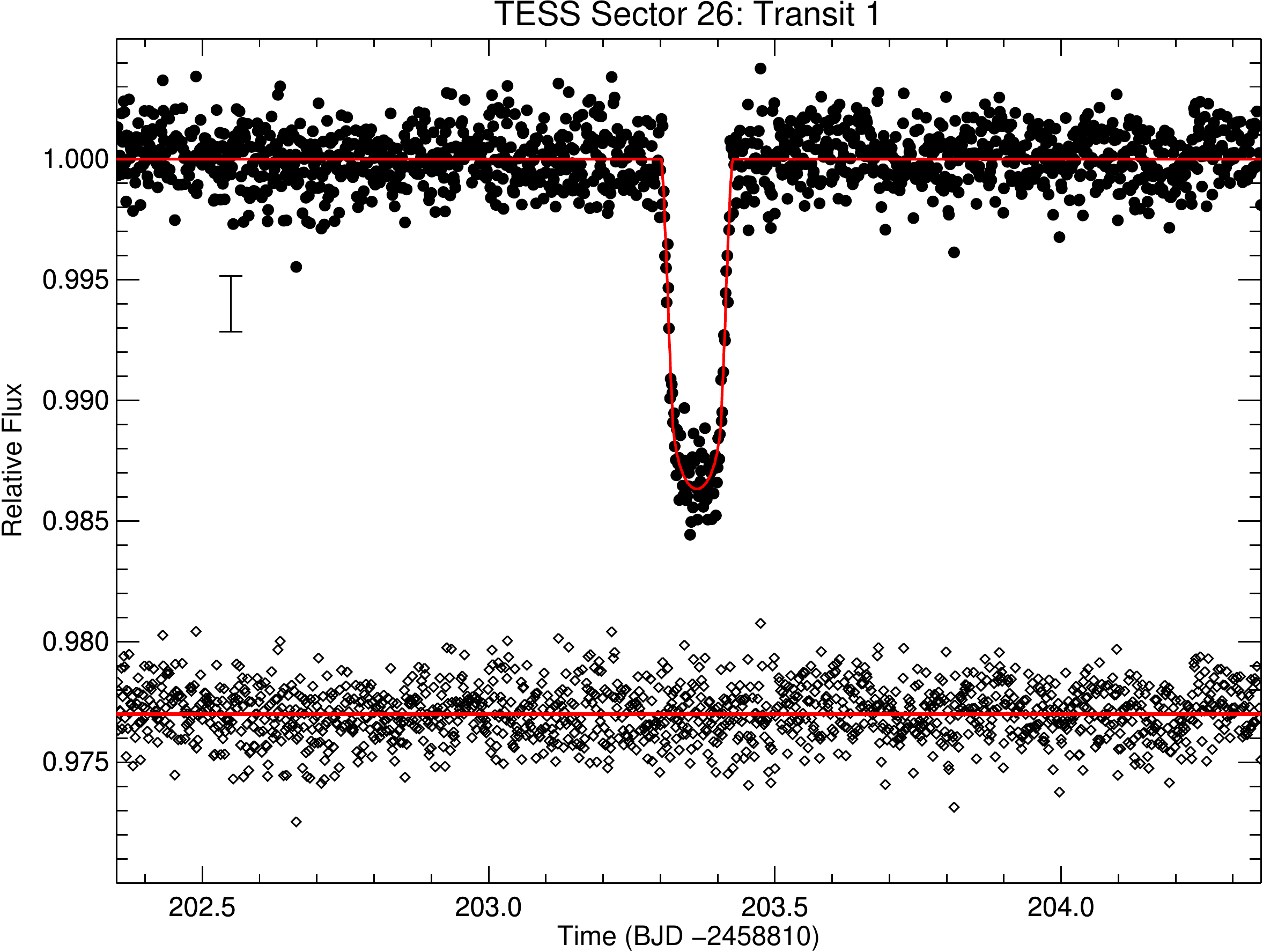} & 
  \includegraphics[width=0.50\textwidth]{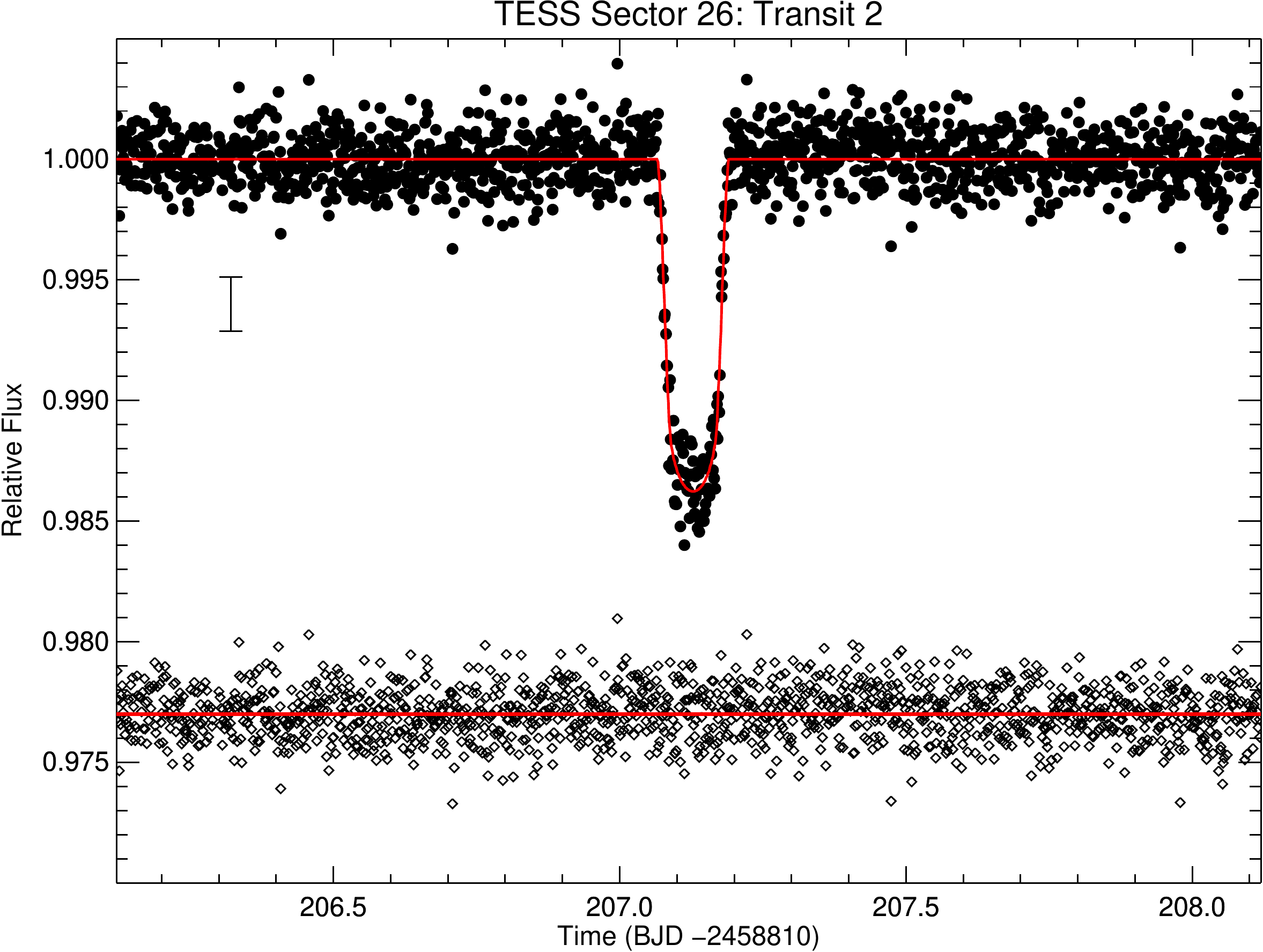} \\
    \includegraphics[width=0.50\textwidth]{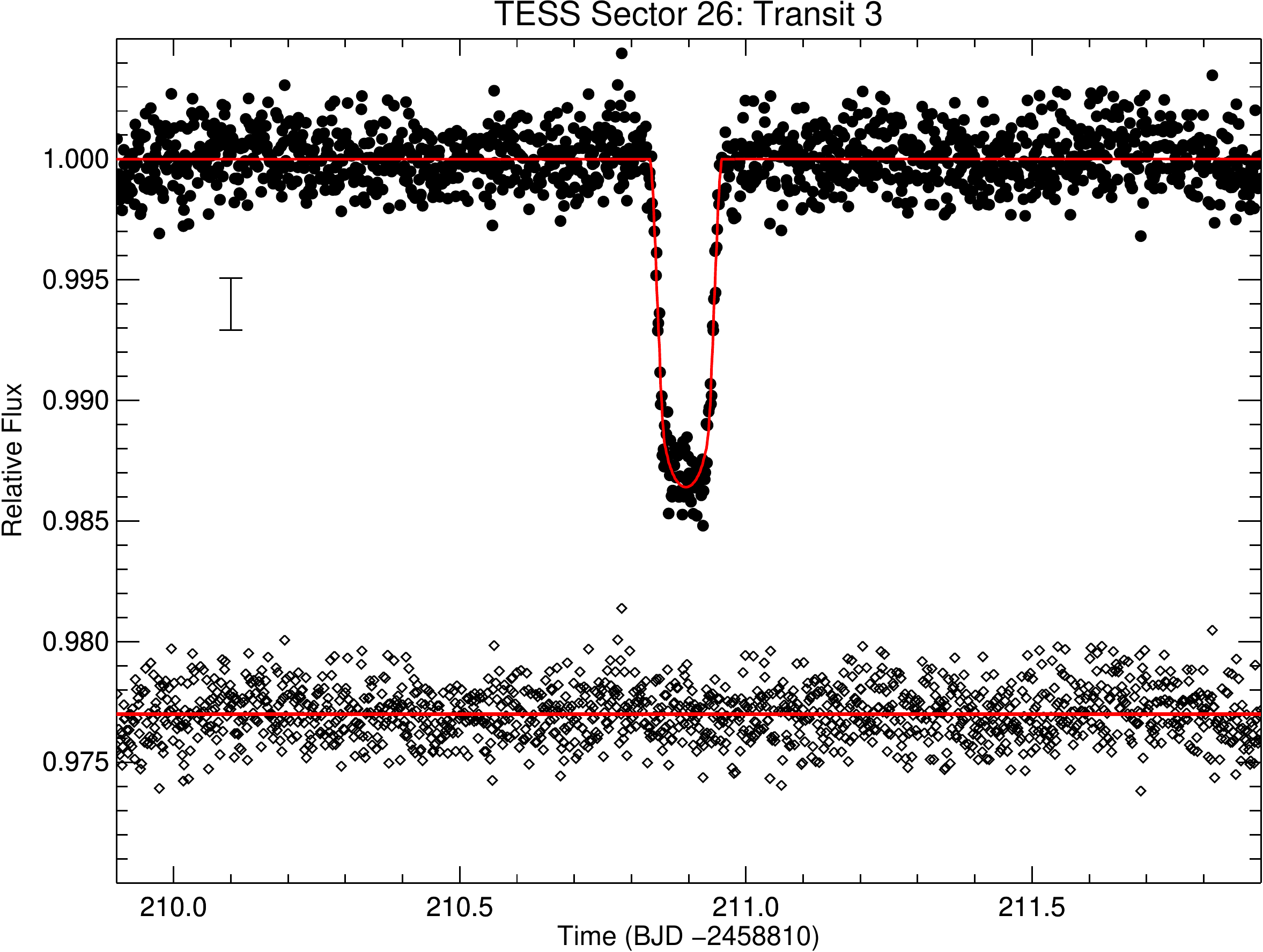} &
 \includegraphics[width=0.50\textwidth]{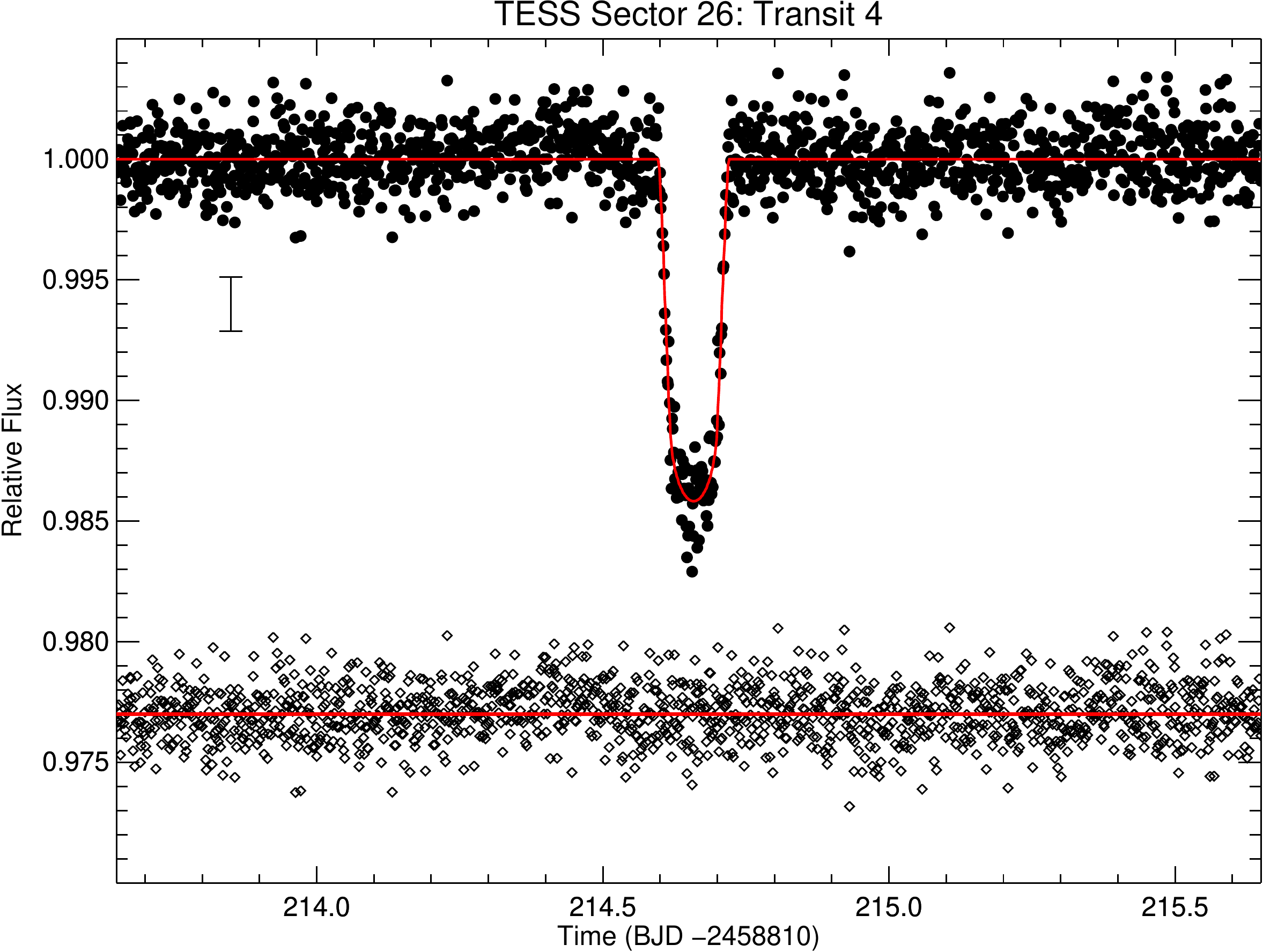} \\
  \includegraphics[width=0.50\textwidth]{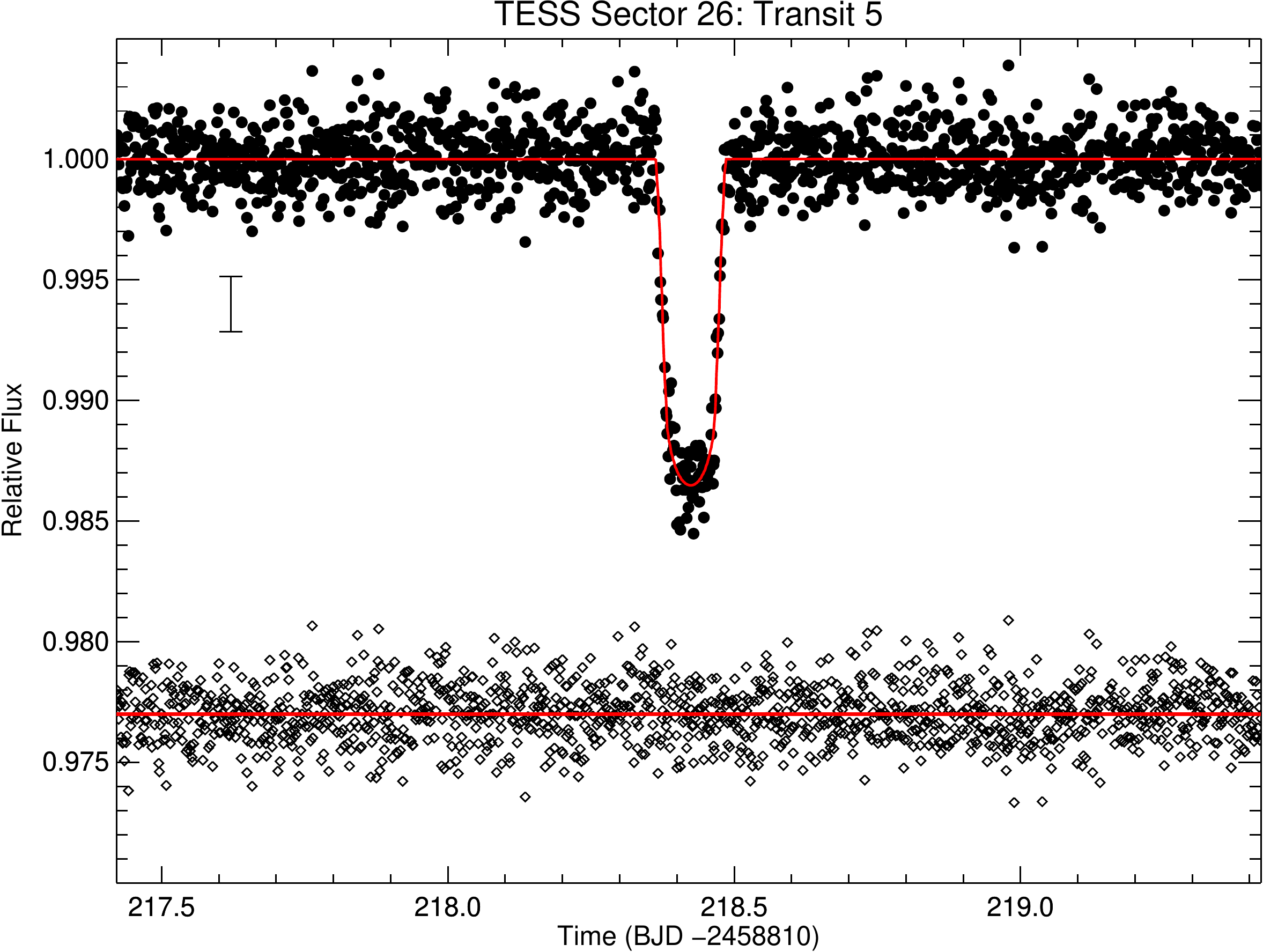} &
    \includegraphics[width=0.50\textwidth]{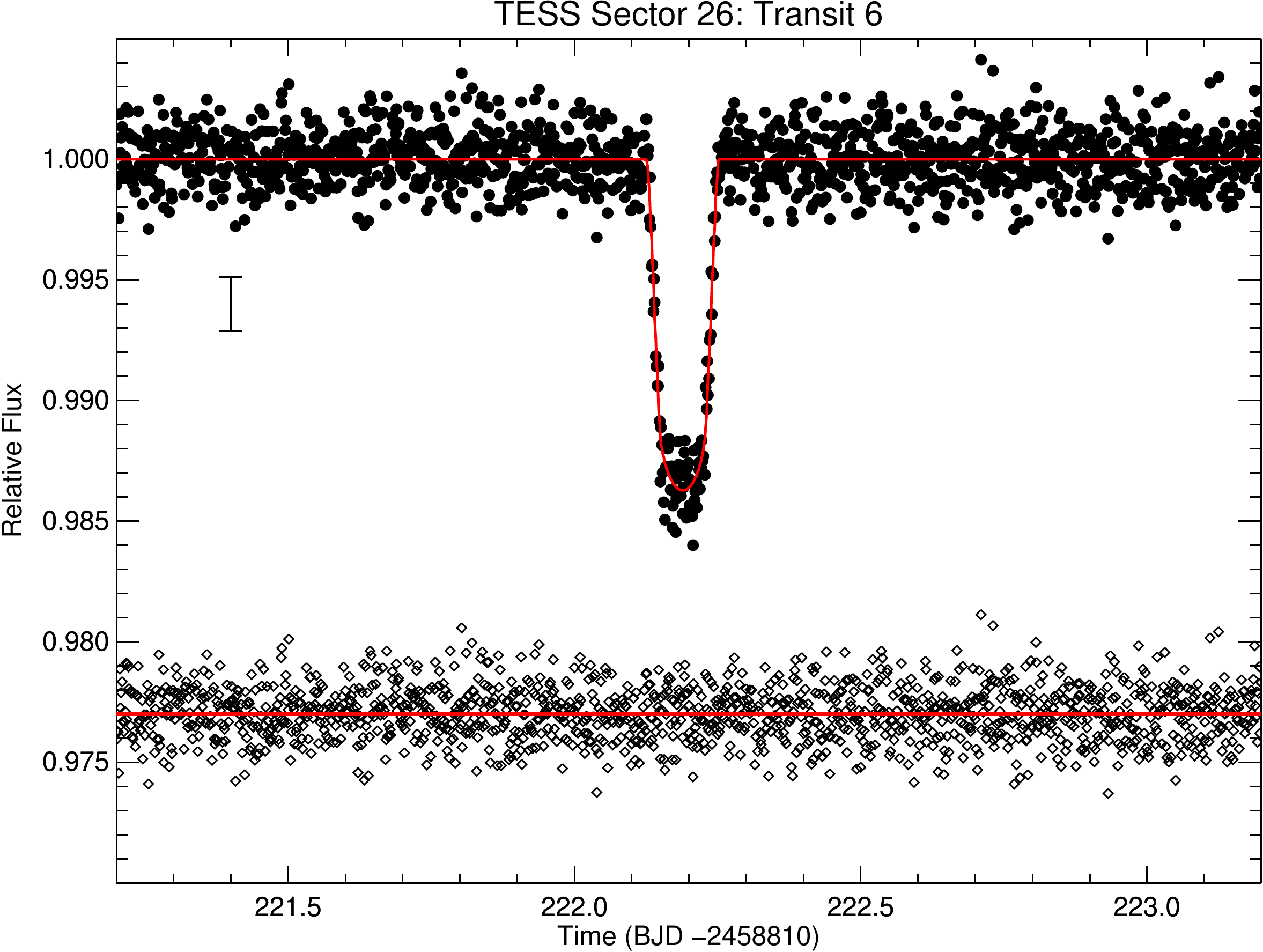} \\
  \end{tabular}
\caption{Individual TESS transit events from Sector 26 of XO-6b. The best-fitting model obtained from the EXOplanet MOdeling Package (\texttt{EXOMOP}) is shown as a solid red line. The residuals (light curve - model) are shown below the light curve.}
\label{fig:ind_transits_sec26}
\end{figure*}

\newpage
\clearpage
\section{Parameters describing the light-time effect (LiTE) model of Garai et al. (2020)}\label{app:Garai2020_model}

\begin{table}[htb]
\caption{Parameters describing the light-time effect (LiTE) model of \citet{Garai2020} that gave the best fit to their reported transit timing variations.}
    \centering
    \begin{tabular}{lcc}
     \hline
     Parameter &  units    & value   \\ \hline
     $P_{orb3}$ &  days    & 456   \\
     $a \sin(i_3)$ &    & 2.03   \\
     $e_3$ &     & 0.85   \\
     $T(0)_3  $ &  BJD    & 	2458184   \\
    $\omega_3$ &  \degree    & 		53   \\
    $K_3$ &  mins    & 		14.6   \\
    $f(M_3)$ &  M$_\odot$    & 	5.3   \\
   \hline
   \multicolumn{3}{l}{Where all parameters refer to the third body.} \\
\multicolumn{3}{l}{$P_{orb3}$ is the orbital period,} \\
\multicolumn{3}{l}{$a$ is the semi-major axis,} \\
\multicolumn{3}{l}{$i_3$ the inclination,} \\
\multicolumn{3}{l}{$e_3$ is the eccentricity,} \\ 
\multicolumn{3}{l}{$T(0)_3$ is the pericenter passage time,} \\
\multicolumn{3}{l}{$\omega_3$ is the pericenter longitude,} \\
\multicolumn{3}{l}{$K_3$ is the semi-amplitude,} \\
\multicolumn{3}{l}{$f(M_3)$ is the mass function, $(a \sin(i_3))^3/P^2_{orb3}$.} \\
    \end{tabular}
    \label{tb:Garai_LiTE_model_params}
\end{table}

\newpage
\clearpage
\section{O-C values from Garai et al. (2020) and barycentric corrections at corresponding transit epochs}\label{app:barycor_plot}

\begin{figure*}[h!b]
\plotone{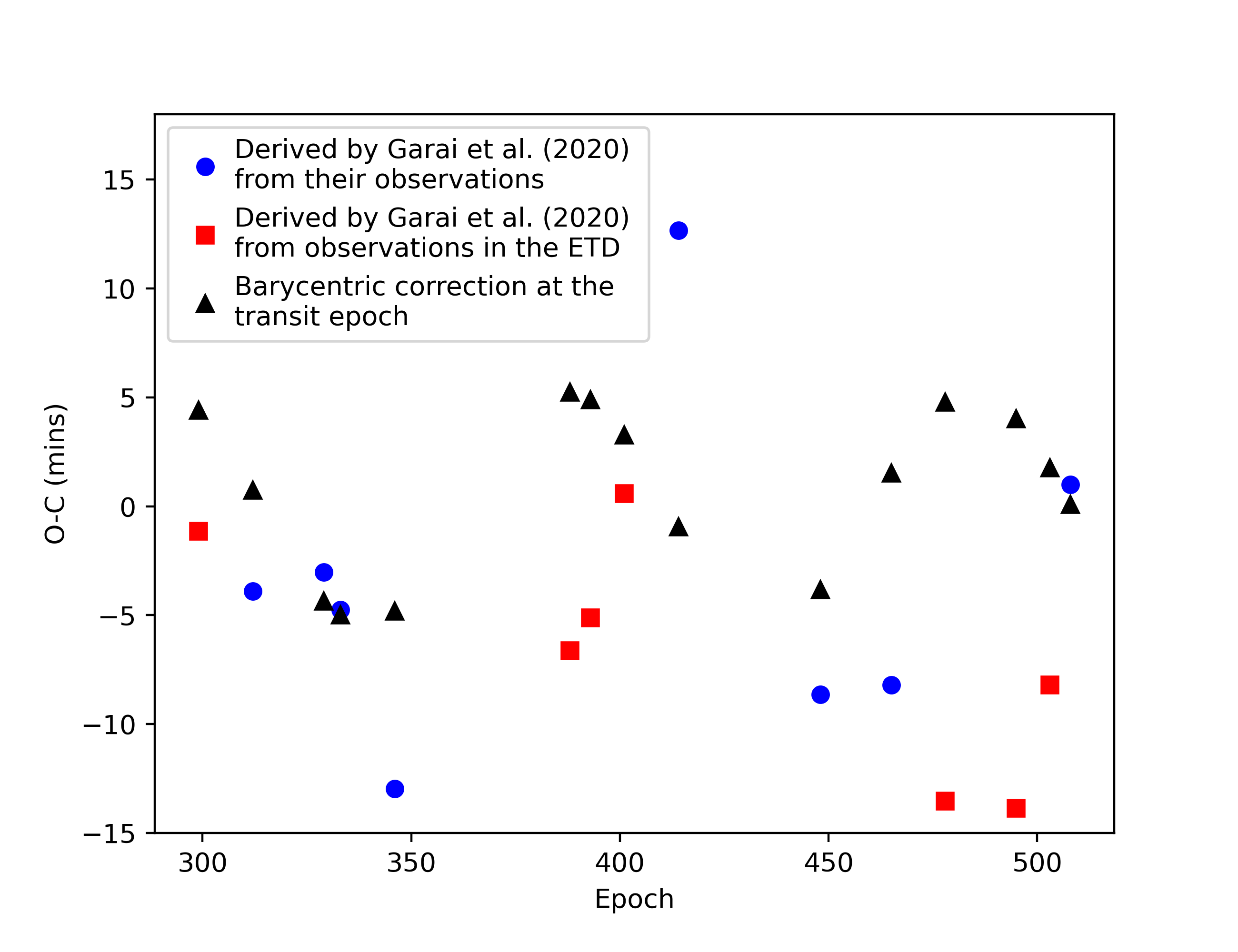}
\caption{The TTVs reported by \citet{Garai2020} from their own observations (blue circles) and observations in the ETD (red squares).  Additionally, the black triangles show the barycentric correction required at each transit epoch.}
\label{fig:barycor}
\end{figure*}


\bibliography{references}{}
\bibliographystyle{aasjournal}



\end{document}